\documentclass{aa}
\usepackage{graphicx}
\usepackage{multirow}
\usepackage{txfonts}
\usepackage[]{hyperref}

\begin{document} 

\title{The Hubble PanCET program: An extensive search for metallic ions in the exosphere of GJ~436~b}

\author{L.~A.~dos~Santos
\inst{1}
\and
D.~Ehrenreich
\inst{1}
\and
V.~Bourrier
\inst{1}
\and
A.~Lecavelier~des~Etangs
\inst{2}
\and
M.~L\'opez-Morales
\inst{3}
\and
D.~K.~Sing
\inst{4, 5}
\and
G.~Ballester
\inst{6}
\and
L.~Ben-Jaffel
\inst{2}
\and
L.~A.~Buchhave
\inst{7}
\and
A.~Garc\'ia~Mu\~noz
\inst{8}
\and
G.~W.~Henry
\inst{9}
\and
T.~Kataria
\inst{10}
\and
B.~Lavie
\inst{1}
\and
P.~Lavvas
\inst{11}
\and
N.~K.~Lewis
\inst{12}
\and
T.~Mikal-Evans
\inst{13}
\and
J.~Sanz-Forcada
\inst{14}
\and
H.~Wakeford
\inst{15}
}

\institute{Observatoire astronomique de l’Université de Genève, 51 
chemin des Maillettes, 1290 Versoix, Switzerland\\
\email{Leonardo.dosSantos@unige.ch}
\and
Institut d'Astrophysique de Paris, CNRS, UMR 7095 \& Sorbonne Universités
UPMC Paris 6, 98 bis bd Arago, 75014 Paris, France
\and
Center for Astrophysics | Harvard \& Smithsonian, 60 Garden Street, Cambridge, MA 02138, USA
\and
Department of Earth \& Planetary Sciences, Johns Hopkins University, Baltimore, MD, USA
\and
Department of Physics \& Astronomy, Johns Hopkins University, Baltimore, MD, USA
\and
Lunar and Planetary Laboratory, University of Arizona, Tucson, AZ 85721, USA
\and
DTU Space, National Space Institute, Technical University of Denmark, Elektrovej 328, DK-2800 Kgs. Lyngby, Denmark
\and
Zentrum f\"ur Astronomie und Astrophysik, Technische Universit\"at Berlin, Hardenbergstrasse 36, D-10623 Berlin, Germany
\and
Center of Excellence in Information Systems, Tennessee State University, Nashville, TN 37209, USA
\and
NASA Jet Propulsion Laboratory, 4800 Oak Grove Drive, Pasadena, CA 91109, USA
\and
Groupe de Spectrom\'etrie Mol\'eculaire et Atmosph\'erique, Universit\'e de Reims, Champagne-Ardenne, CNRS UMR F-7331, France
\and
Department of Astronomy and Carl Sagan Institute, Cornell University, 122 Sciences Drive, 14853, Ithaca, NY, USA
\and
Kavli Institute for Astrophysics and Space Research, Massachusetts Institute of Technology, 77 Massachusetts Avenue, 37-241, Cambridge, MA 02139, USA
\and
Centro de Astrobiolog\'{i}a (CSIC-INTA), E-28692 Villanueva de la Ca\~nada, Madrid, Spain
\and
Space Telescope Science Institute, 3700 San Martin Drive, Baltimore, MD 21218, USA
}

\date{Received 11 April 2019 / Accepted 10 July 2019}

\abstract
{The quiet M2.5 star GJ 436 hosts a warm Neptune that displays an extended atmosphere that dwarfs its own host star. Predictions of atmospheric escape in such planets state that H atoms escape from the upper atmosphere in a collisional regime and that the flow can drag heavier atoms to the upper atmosphere. It is unclear, however, what astrophysical mechanisms drive the process.}
{Our objective is to leverage the extensive coverage of observations of the far-ultraviolet (FUV) spectrum of GJ~436 obtained with the Cosmic Origins Spectrograph (COS) to search for signals of metallic ions in the upper atmosphere of GJ~436~b, as well as study the activity-induced variability of the star.}
{We analyzed flux time-series of species present in the FUV spectrum of GJ~436 and successfully performed geocoronal contamination removal in the COS Lyman-$\alpha$ profiles obtained near the Earth's night-side.}
{GJ~436 displays flaring events with a rate of $\sim$10 d$^{-1}$. There is evidence for a possibly long-lived active region or longitude that modulates the FUV metallic lines of the star with amplitudes up to 20\%. Despite the strong geocoronal contamination in the COS spectra, we detected in-transit excess absorption signals of $\sim$50\% and $\sim$30\% in the blue and red wings, respectively, of the Lyman-$\alpha$ line. We rule out a wide range of excess absorption levels in the metallic lines of the star during transit.}
{The large atmospheric loss of GJ~436~b observed in Lyman-$\alpha$ transmission spectra is stable over the timescale of a few years, and the red wing signal supports the presence of a variable hydrogen absorption source besides the stable exosphere. The previously claimed in-transit absorption in the \ion{Si}{III} line is likely an artifact resulting from the stellar magnetic cycle. The non-detection of metallic ions in absorption could indicate that the escape is not hydrodynamic or that the atmospheric mixing is not efficient in dragging metals high enough for sublimation to produce a detectable escape rate of ions to the exosphere.}


\titlerunning{An extensive search for metallic ions in the exosphere of GJ~436~b}

\maketitle
%

\section{Introduction}

Atmospheric escape is an important process that dictates planetary evolution and habitability in the Solar System \citep[e.g.,][]{1987Icar...71..203P, 2003Icar..165....9L, 2006P&SS...54.1425K} and extra-solar systems \citep[e.g.,][]{2011A&A...532A...6S, 2015Icar..250..357C, 2017ApJ...847L...4D, 2017MNRAS.464.3728B}. Hydrodynamic escape in strongly irradiated planets is driven by extreme ultraviolet (XUV) irradiation from their host stars \citep{2003Natur.422..143V, 2003ApJ...598L.121L, 2007P&SS...55.1426G, 2012MNRAS.425.2931O}, which is presumed to be strongest in the early history of a given planetary system \citep{2005ApJ...622..680R, 2007LRSP....4....3G}. Further evidence for the importance of atmospheric escape in planetary evolution comes from studies of transiting exoplanet populations, which brought our attention to a dearth of short-period ($P < 10$ d) planets with masses between 0.01 and 1 M$_\mathrm{J}$ \citep[the hot Neptune desert;][]{2007A&A...461.1185L, 2009MNRAS.396.1012D, 2011A&A...529A.136E, 2011ApJ...727L..44S, 2012ApJ...761...59L, 2016A&A...589A..75M, 2018MNRAS.476.5639I}. Despite the detection of many ultra short-period small (presumably rocky) planets, the \emph{Kepler} satellite found almost no strongly irradiated Neptune-size planets, even though the survey was much more sensitive to larger planets \citep{2012ApJS..201...15H}.

Transiting exoplanets amenable to atmospheric characterization offer one of the most compelling opportunities to study planetary evolution. The first observation of Na in the optical transmission spectrum of the giant exoplanet HD~209458~b \citep{2002ApJ...568..377C} was a crucial milestone toward this goal, generating several theoretical and experimental efforts to probe exoplanetary atmospheres. In particular, the first direct evidence of atmospheric escape in exoplanets was reported for the hot Jupiters HD~209458~b \citep{2003Natur.422..143V, 2004ApJ...604L..69V, 2008A&A...483..933E, 2015ApJ...804..116B} and HD~189733~b \citep{2010A&A...514A..72L, 2012A&A...543L...4L, 2013A&A...551A..63B, 2013A&A...553A..52B}, using far-ultraviolet (FUV) transit spectroscopy with the Hubble Space Telescope (\emph{HST}). However, to this date, the most spectacular observation of atmospheric escape remains that of GJ~436~b. It displays a transit depth of 56\% and a long egress in the blue wing of the Lyman-$\alpha$ line caused by the extended tail of neutral hydrogen that escapes vigorously from the planet \citep{2014ApJ...786..132K, 2015Natur.522..459E, 2015A&A...582A..65B, 2016A&A...591A.121B, 2017A&A...605L...7L}.

GJ~436~b is a warm Neptune exoplanet orbiting a nearby and relatively quiet M2.5 dwarf \citep{2004ApJ...617..580B, 2007A&A...472L..13G}. The planet lies in the lower-mass edge of the hot Neptune desert (see the stellar and planetary parameters in Table \ref{gj436_param}). One of the most important mechanisms to explain the hot Neptune desert is the erosion of inflated envelopes rich in H and He, and the observation of large-scale atmospheric escape from GJ~436~b seems to corroborate this hypothesis. According to \citet{2016A&A...591A.121B}, the current atmospheric loss rate of GJ~436~b is $\sim$$1/18000$ Gyr$^{-1}$ in planetary mass fraction, which is not large enough to carve the hot Neptune desert. Moreover, the eccentricity and orbital misalignment of the planet with the spin of the star suggests that it may have recently migrated inward due to an undetected outer companion as of yet \citep{2012A&A...545A..88B, 2014ApJ...796...32S, 2018Natur.553..477B}. In contrast, \citet{2018A&A...620A.147B} showed that the warm Neptune GJ~3470~b displays a large mass loss rate comparable to that of hot Jupiters, rendering it the most extreme case of mass loss observed to date. GJ~3470~b could already have lost up to 40\% of its mass over its 2 Gyr lifetime, suggesting that planetary mass loss has the potential to change the population of close-in giant exoplanets.

Several questions about the GJ~436 system remain unanswered, such as: how stellar activity affects the stellar FUV energy output, whether the planet loses other species besides hydrogen, and whether we should expect other warm Neptunes around M dwarfs to display similar escape rates \citep[e.g.,][]{2018A&A...620A.147B}. We note that FUV transit spectra with \emph{HST} will help answer these questions, but several datasets may be necessary since the stellar lines of GJ~436 are weak when compared to solar-type stars. In particular, the Cosmic Origins Spectrograph \citep[COS;][]{2012ApJ...744...60G} has a wider wavelength range and is more sensitive than the Space Telescope Imaging Spectrograph \citep[STIS;][]{1998PASP..110.1183W} in the FUV, giving us access to several metallic stellar lines.

\citet{2017A&A...605L...7L} reported a tentative absorption signal in the \ion{Si}{III} stellar line (1206.5 \AA) of GJ~436 that could be of a planetary nature. If proven to be accurate, this signal would suggest that Si atoms are hydrodynamically dragged from the lower atmosphere of the planet by the H atoms \citep[similarly to HD~209458~b; ][]{2003Natur.422..143V, 2010ApJ...723..116K}, indicating the presence of atmospheric mixing and clouds in the lower atmosphere \citep{2010ApJ...716.1060V}, which is consistent with the flat infrared-optical transit spectrum of the planet \citep{2014Natur.505...66K, 2018AJ....155...66L}. In contrast, \citet{2017ApJ...834L..17L} reported a non-detection of \ion{C}{II} and \ion{Si}{III} in-transit absorption signals in two \emph{HST} visits during the transit of GJ~436~b, assuming a large asymmetric transit light curve as seen in Lyman-$\alpha$. Their simulations predict \ion{C}{II} transit depths of 2\% and 19\% in the full line passband and line center, respectively. 

We report here on the analysis of several \emph{HST}-COS observations covering different phases of the planetary transit in four epochs, aiming to resolve the questions about GJ~436~b and the hydrodynamical nature of the atmospheric escape process. This manuscript has the following structure: in Sect. \ref{obs} we describe the observations and the post-processing necessary after data reduction; in Sects. \ref{flares} and \ref{rot_modulation_sect} we examine the impact of activity (flares and rotational modulation) in the FUV fluxes of GJ~436; in Sect. \ref{lya_cos} we present the first detection of the deep Lyman-$\alpha$ transit of GJ~436~b using \emph{HST}-COS; in Sect. \ref{results} we discuss the results of the search for metallic ions in the exosphere of GJ~436~b; and in Sect. \ref{conclusions} we summarize our conclusions and present future research perspectives for GJ~436~b.

\section{Observations and data reduction}\label{obs}

GJ~436 b is one of the targets of the Hubble Panchromatic Comparative Exoplanet Treasury (PanCET) program GO-14767 \citep[PIs: D. Sing and M. L\'opez-Morales; see][]{2017ApJ...835L..12W, 2017Natur.548...58E, 2018MNRAS.474.1705N, 2018AJ....156..298A, 2018A&A...620A.147B}. GJ~436 was observed during four visits using the Cosmic Origins Spectrograph (COS) fed by the Hubble Space Telescope, using the grating G130M centered on 1291 \AA. These visits were planned to include at least one orbit during the optical primary transit of the planet GJ~436~b, according to the ephemeris of \citet{2014A&A...572A..73L}. In total, two orbits, one of them while in-transit, were affected by technical failures and did not register counts since the shutter of the instrument was closed; these orbits are discarded from the analysis. In this study we also made use of \emph{HST}-COS archival observations obtained during programs GO-15174 (PI: R. O. Loyd) and GO-13650 (MUSCLES Treasury Survey; PI: K. France). The following observations log is located in Table \ref{log}: for visits A-D (PanCET program), we have a total of 18 usable orbits of which three are in transit; in visits E and F (MUSCLES program) there are eight orbits and none of them are in-transit; the eight single-orbit visits of program GO-15174 were meant to cover a wide swath of phases of the orbit of GJ~436~b and do not cover the transit.

\begin{figure*}
\centering
\includegraphics[width=0.9\hsize]{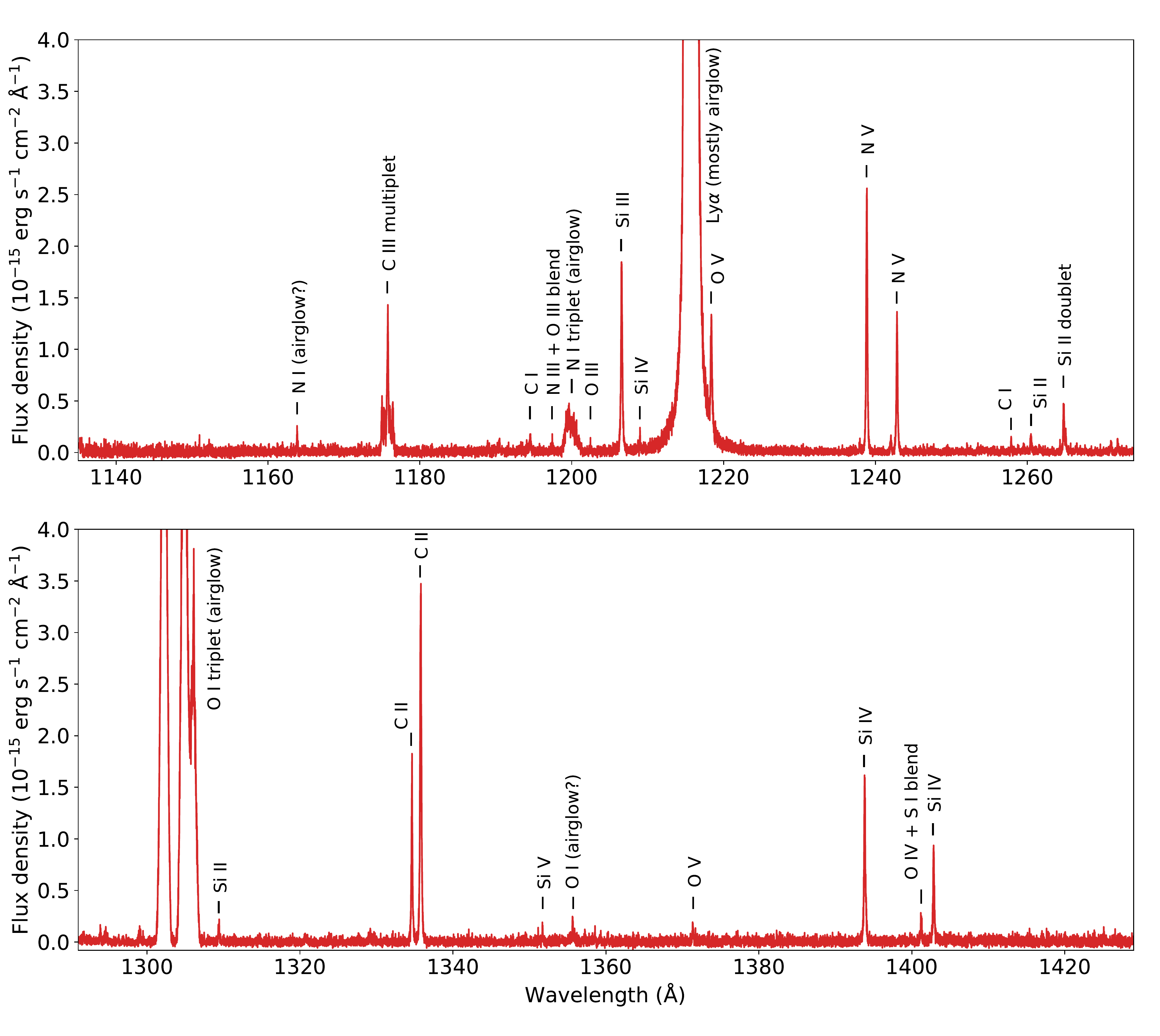}
\caption{Combined FUV spectrum of GJ~436 measured with \emph{HST}-COS using G130M grating centered on 1291 \AA.}
\label{GJ436_spectrum}
\end{figure*}

The raw spectra were processed automatically by the instrument's pipeline. Since the observations were performed in time-tag mode, we were able to split the data in sub-exposures using the \texttt{calcos} package from the AstroConda software stack\footnote{\footnotesize{Available at \url{http://astroconda.readthedocs.io}.}}. We performed our analysis using the same FUV line list as in Table \ref{line_list}. For a reference, we combined all the \emph{HST}-COS observations of GJ~436 that are publicly available and produced a high signal-to-noise FUV spectrum, which we reproduce in Fig. \ref{GJ436_spectrum}.

\begin{table}
\caption{Stellar and planetary parameters of GJ~436 and GJ~436~b.}
\label{gj436_param}
\centering
\begin{tabular}{l c c}
\hline\hline
\multicolumn{2}{c}{Stellar parameters of GJ~436} & Ref. \\
\hline
Radius & $0.449 \pm 0.019$ R$_\odot$ & (a) \\
Mass & $0.445 \pm 0.044$ M$_\odot$ & (a) \\
Eff. temperature & $3479 \pm 60$ K & (a) \\
Proj. rot. velocity & $0.330^{+0.091}_{-0.066}$ km s$^{-1}$ & (b) \\
Rotational period & $44.09 \pm 0.08$ d & (b) \\
Inclination rot. axis & $39^{+13}_{-9} \deg$ & (b) \\
Distance & $9.756 \pm 0.009$ pc & (c) \\
$\log{R'_\mathrm{HK}}$ & $-5.32 \pm 0.07$ & (d) \\
$L_\mathrm{X} / L_\mathrm{Bol}$ & $1.950 \times 10^{-6}$ & (e) \\
\hline
\multicolumn{2}{c}{Planetary parameters of GJ~436~b} & Ref. \\
\hline
Radius & $4.04 \pm 0.85$ R$_\oplus$ & (f) \\
Mass & $25.4^{+2.1}_{-2.0}$ M$_\oplus$ & (f) \\
Orbital period & $2.64389803 \pm 0.00000026$ d & (f) \\
Semi-major axis & $14.54 \pm 0.14$ R$_\star$ & (f) \\
Ref. time (BJD) & $2454865.084034 \pm 0.000035$ & (f) \\
Orbital inclination & $88.858^{+0.049}_{-0.052} \deg$ & (b) \\
Eccentricity & $0.1616 \pm 0.004$ & (f) \\
Arg. periastron & $327.2^{+1.8}_{-2.2} \deg$ & (f) \\
\hline
\end{tabular}
\tablebib{(a) \citet{2015ApJ...804...64M}, (b) \citet{2018Natur.553..477B}, (c) \citet{2018A&A...616A...1G}, (d) \citet{2015MNRAS.452.2745S}, (e) \citet{2011A&A...532A...6S}, (f) \citet{2014A&A...572A..73L}.}
\end{table}

\begin{table}
\caption{Observations log of GJ~436 with \emph{HST}-COS centered at 1291~\AA.}
\label{log}
\centering
\begin{tabular}{l c c c c}
\hline\hline
\multirow{2}{*}{Visit} & \multirow{2}{*}{Orbit} & Start time & Exp. time & Phase \\
& & (UT) & (s) & (h) \\
\hline
\multicolumn{5}{c}{Hubble PanCET program} \\
\hline
\multirow{5}{*}{A} & 1 & 2017-11-19 20:30:21 & 1881.184 & -1.32 \\
& 2\tablefootmark{$\dagger$} & 2017-11-19 21:49:34 & 2702.08 & $\cdots$ \\
& 3 & 2017-11-19 23:24:55 & 2702.176 & +1.71 \\
& 4 & 2017-11-20 01:00:15 & 2702.112 & +3.30 \\
& 5 & 2017-11-20 02:36:06 & 2702.176 & +4.89 \\
\hline
\multirow{5}{*}{B} & 1 & 2017-12-21 12:17:30 & 1881.152 & -2.97 \\
& 2 & 2017-12-21 13:34:53 & 2702.144 & -1.57 \\
& 3 & 2017-12-21 15:10:16 & 2702.144 & +0.02 \\
& 4 & 2017-12-21 16:45:38 & 2702.144 & +1.61 \\
& 5 & 2017-12-21 18:21:00 & 2702.144 & +3.20 \\
\hline
\multirow{5}{*}{C} & 1 & 2018-01-24 21:02:28 & 1688.192 & -3.15 \\
& 2 & 2018-01-24 22:17:19 & 2702.176 & -1.76 \\
& 3 & 2018-01-24 23:52:39 & 2702.176 & -0.17 \\
& 4 & 2018-01-25 01:27:59 & 2702.176 & +1.42 \\
& 5 & 2018-01-25 03:03:19 & 2702.176 & +3.01 \\
\hline
\multirow{5}{*}{D} & 1 & 2018-02-28 07:32:12 & 1688.096 & -1.55 \\
& 2 & 2018-02-28 08:48:36 & 2702.176 & -0.13 \\
& 3 & 2018-02-28 10:23:56 & 2702.176 & +1.46 \\
& 4\tablefootmark{$\dagger$} & 2018-02-28 11:59:18 & 2702.176 & $\cdots$ \\
& 5 & 2018-02-28 13:34:39 & 2702.176 & +4.64 \\
\hline
\multicolumn{5}{c}{MUSCLES program} \\
\hline
\multirow{3}{*}{E}& 1 & 2012-06-23 07:22:56 & 980.192 & -14.76 \\
& 2 & 2012-06-23 07:41:15 & 1191.168 & -14.43 \\
& 3 & 2012-06-23 08:47:20 & 1200.192 & -13.32\\
\hline
\multirow{5}{*}{F}& 1 & 2015-06-25 23:37:36 & 1243.168 & -3.68 \\
& 2 & 2015-06-26 00:43:54 & 2713.184 & -2.37 \\
& 3 & 2015-06-26 02:19:20 & 2713.216 & -0.80 \\
& 4 & 2015-06-26 03:54:46 & 2713.184 & +0.79 \\
& 5 & 2015-06-26 05:30:12 & 2713.216 & +2.38 \\
\hline
\multicolumn{5}{c}{Program GO-15174} \\
\hline
\multirow{8}{*}{$\cdots$} & 1 & 2017-12-22 21:35:19 & 1957.152 & +30.34 \\
& 2 & 2017-12-23 21:25:43 & 1957.184 & -9.28 \\
& 3 & 2017-12-24 09:55:32 & 1957.152 & +3.22 \\
& 4 & 2017-12-24 16:29:58 & 1957.152 & +9.79 \\
& 5 & 2018-01-19 10:24:56 & 1850.176 & -6.84 \\
& 6 & 2018-01-19 15:27:48 & 1850.144 & -1.79 \\
& 7 & 2018-02-08 13:52:25 & 1850.144 & -31.01 \\
& 8 & 2018-02-23 08:19:12 & 1850.144 & +6.17 \\
\hline
\end{tabular}
\tablefoot{Phases are in relation to the orbit of GJ~436~b. Orbits marked with \tablefoottext{$\dagger$}\ had a pointing failure and did not register counts; these orbits were discarded from our analysis. Observations of program GO-15174 were performed with single-orbit visits.}
\end{table}

We found that the spectral lines of GJ~436 are systematically shifted from the stellar rest frame in our datasets, displaying excess Doppler shifts from -6 to 20 km s$^{-1}$ over the systemic velocity of the star \citep[$v_\mathrm{R} = 9.61$ km s$^{-1}$;][]{2002ApJS..141..503N}. The variation occurs at different levels depending on the position of the line in the detector and the time of the observation. However, orbits from the same visit tend to display similar line-to-line Doppler shifts. This wavelength calibration issue has been previously observed with COS \citep[e.g.,][]{2012ApJ...754...69L, 2017ApJ...834L..17L, 2018A&A...615A.117B}.

In order to correct for these systematic Doppler shifts, we measured them in a line-by-line and visit-by-visit fashion for all lines in the spectra, except for the lines contaminated by airglow (see Sect. \ref{ag_corr}) and those that are too faint. In the latter case, we consider that Doppler shifts changed by the same value as the closest line that we could measure. The Doppler shifts are quantified as an average per visit by fitting Gaussian profiles (whose parameters are fit at the position of the Gaussian and its amplitude) to the stellar lines. The correction for Doppler velocity shifts is applied during the computation of fluxes for each line in each spectrum. For each spectral line, the fluxes are computed by integrating the flux densities in their corresponding passbands (see Table \ref{line_list}). When applicable, we accumulated the fluxes of multiplets to improve the signal-to-noise ratio of the emission for each species. In addition, we removed the subexposures with possible flaring activity (Sect. \ref{flares}).

\begin{table}
\caption{Spectral line list used in this work.}
\label{line_list}
\centering
\begin{tabular}{l c c}
\hline\hline
Ion & Central wavelength ($\mathrm{\AA}$) & Integration range (km s$^{-1}$) \\
\hline
\ion{C}{III} & 1175.59 & [-240, +230] (multiplet) \\
\ion{Si}{III} & 1206.5 & [-50, +50] \\
\ion{H}{I} & 1215.6702 & See Fig. \ref{lya_pr} \\
\ion{O}{V} & 1218.344 & [-50, +50] \\
\ion{Si}{II} & 1264.738 & [-50, +100] (doublet) \\
\hline
\multirow{2}{*}{\ion{N}{V}} & 1238.821 & [-80, +80] \\
& 1242.804 & [-70, +70] \\
\hline
\multirow{2}{*}{\ion{C}{II}} & 1334.532 & [-50, +50] \\
& 1335.708 & [-60, +60] \\
\hline
\multirow{2}{*}{\ion{Si}{IV}} & 1393.755 & [-50, +50] \\
& 1402.77 & [-40, +40] \\
\hline
\end{tabular}
\tablefoot{More information about the formation of FUV lines can be found in, e.g., \citet{2008ApJS..175..229A}.}
\end{table}

\citet{2017A&A...599A..75W} found that the uncertainties of the spectra processed by the instrument's pipeline are overestimated, especially for fluxes above $10^{-14}$ erg s$^{-1}$ cm$^{-2}$ \AA$^{-1}$. For our analysis, we defined the uncertainties of the spectra according to Eqs. 1 and 2 of \citet{2017A&A...599A..75W}. Furthermore, errors in measuring stellar spectra affect how accurately we measured the flux. In theory, it should be possible to correct these errors if we know their sources and how they affect the data. In particular, systematic instrumental effects of \emph{HST} observations, such errors in flux calibration, and the thermal "breathing" effect \citep[e.g.,][]{2012A&A...547A..18E, 2017A&A...605L...7L} can occur as flux variations resulting from changes in focus that correlate with the orbit of the telescope. We searched for such correlations in the available datasets for GJ~436 and for 55~Cnc~e (V. Bourrier, private communication), but we did not find evidence for significant thermal breathing with \emph{HST}-COS. This is likely because COS has a circular aperture and is thus less sensitive to losses due to focus variations when compared to slit or grism spectrographs, such as STIS and WFC3.

Effects intrinsic to the target being observed that are not known or not taken into account are another important source of errors. One example is the Lyman-$\alpha$ transit of GJ~436~b in which the duration of the transit in optical wavelengths is only one hour \citep{2014A&A...572A..73L}. However, in Lyman-$\alpha$ the transit event lasts more than 20 hours, owing to the large size and shape of the planet's exosphere \citep{2017A&A...605L...7L}. Were this not known, then we would have erroneously measured the baseline Lyman-$\alpha$ flux of the star during the long transit of the planet's exosphere. Another example of measurement error occurs when stellar activity effects, such as modulation by active regions in the stellar surface, are not taken into account.

\section{Flares of GJ~436}\label{flares}

In this section we mainly discuss the results obtained during Program GO-15174, which were measured more than five hours away from the optical transit in order to avoid any possible planetary signal. Although GJ~436 is a quiet star compared to other M dwarfs \citep[e.g.,][]{2015MNRAS.452.2745S}, we observed strong levels of stellar variability in some of the lines in its FUV spectrum.

\subsection{Identification of flares during exposures}

We used the time-tag information from the raw data to divide each \emph{HST}-COS exposure in four subexposures. We found strong correlations (Pearson-$r > 0.7$) between the out-of-transit fluxes of the lines \ion{Si}{III}, \ion{Si}{IV}, \ion{C}{II,} and \ion{C}{III}. If the spread in fluxes was only due to stochastic uncertainties, then there would not be a correlation between line-by-line flux comparisons. Such flux correlations observed in the out-of-transit spectra only appear when systematics are present, so they must be either of instrumental (see Sect. \ref{obs}) or astrophysical origin. The spectral lines with the lowest level of intrinsic variability are the \ion{N}{V} lines at 1239 and 1243 \AA. 

\begin{figure}
\centering
\includegraphics[width=\hsize]{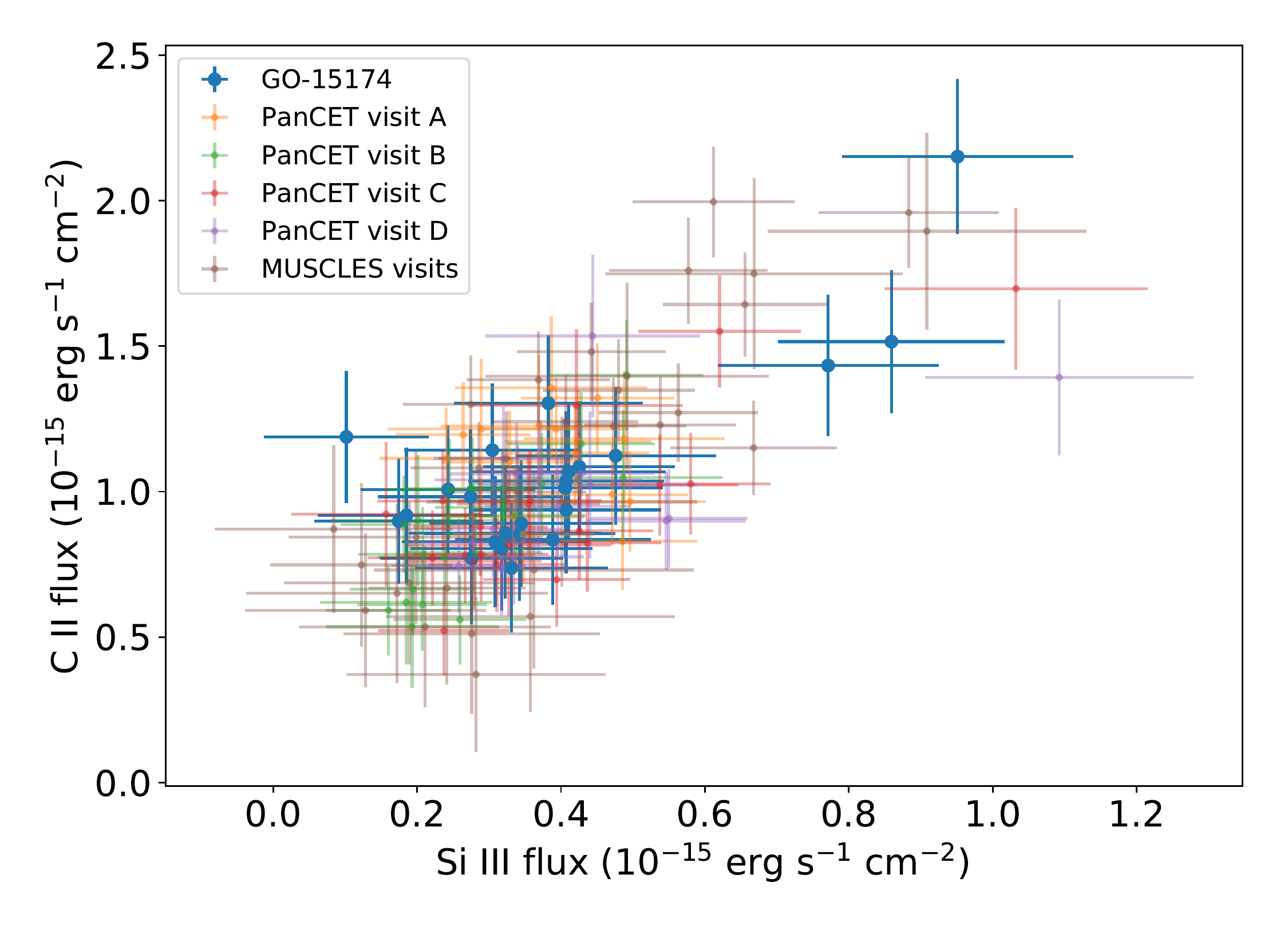}
\caption{Fluxes of \ion{Si}{III} versus \ion{C}{II} lines in spectra of GJ~436. The strong correlation (Pearson-$r > 0.7$) between these fluxes leads us to conclude that the higher-flux end of the plot corresponds to stellar flares, while the lower-flux end corresponds to the quiescent state of the star.}
\label{corr_split}
\end{figure}

Since these flux correlations are wavelength-dependent and being that we did not find evidence for instrumental systematics with \emph{HST}-COS, a sensible first approach is to assume the effect is astrophysical. Following an inspection of the time-tag split light curve of the \ion{Si}{III} and \ion{C}{II} fluxes, we found a statistically significant ($\gtrsim 5 \sigma$) increase in fluxes by 100\% during the first quarter of orbit 2 of program GO-15174; this variation is not seen in the \ion{N}{V} lines. The second half of orbit 8 also displays a similar increase, but with a lower significance. The first quarter of orbit 6 shows an increase in the fluxes of Si lines by 100\% in relation to the average flux of the remaining subexposures, but the same is not seen at high significance in the other lines. We reproduce the \ion{Si}{III} versus \ion{C}{II} line fluxes dispersion map in Fig. \ref{corr_split}, in which each point corresponds to a sub-exposure with \emph{HST}-COS during the out-of-transit program. There are the following two noticeable features in this plot: first, the low fluxes cluster around each other and have a dispersion that is consistent with a weak correlation (probably related to rotational modulation, see Section \ref{rot_modulation_sect}); and second, the higher fluxes are less common and show an apparent correlation. One interpretation for the observed large fluxes is that they result from transient brightenings, while the low-flux end corresponds to the stellar quiescent state. Further, we overplotted the sub-exposure fluxes of the PanCET and MUSCLES visits in Fig. \ref{corr_split} and identified flares in their data as well; these sub-exposures were excised from the data. Orbit 6 of the MUSCLES observations is completely contaminated by a flare, which was originally reported by \citet{2017ApJ...834L..17L}. The flare events in the datasets we analyzed are not limited to a specific phase of the orbit of GJ~436~b. We also observed flares in GJ~436 in X-rays, as well as several other targets in the PanCET survey. The results will be published in a a following article (Sanz-Forcada et al., in prep.); the X-ray light curves of GJ~436 are available publicly in the \texttt{X-exoplanets} database\footnote{\footnotesize{\url{http://sdc.cab.inta-csic.es/xexoplanets/}}}.

\subsection{Discussion}

As an M2.5-type dwarf, GJ~436 is near the limit where stars become fully convective \citep{2011ApJ...743...48W} and start to display strong activity signals. According to \citet{2017ApJ...849...36Y}, 10-15\% of stars of this spectral type display flare behavior; moreover, M dwarfs with a rotational period similar to GJ~436 \citep[44.09 d;][]{2018Natur.553..477B} tend to have flare activity levels near $6 \times 10^{-6}$ L$_\mathrm{flare}$/L$_\mathrm{bol}$ (two orders of magnitude lower than the fastest-rotating M dwarfs in the Kepler field). It is not completely clear, however, if these results can accurately be applied to flaring activity in FUV wavelengths. Previous observations of GJ~436 for the MUSCLES Treasury Survey have also resulted in the detection of flares in \ion{C}{II} and \ion{Si}{III} lines, although they are less frequent and weaker than in other M dwarfs in the program \citep{2017ApJ...843...31Y, 2017ApJ...834L..17L}. Flares with similar levels of brightening are also observed in X-ray light curves of the Sun and have durations between 1 and 20 minutes \citep{1995PASJ...47..251S}. 

\begin{figure*}
    \centering
    \begin{tabular}{cc}
        \includegraphics[width=0.47\hsize]{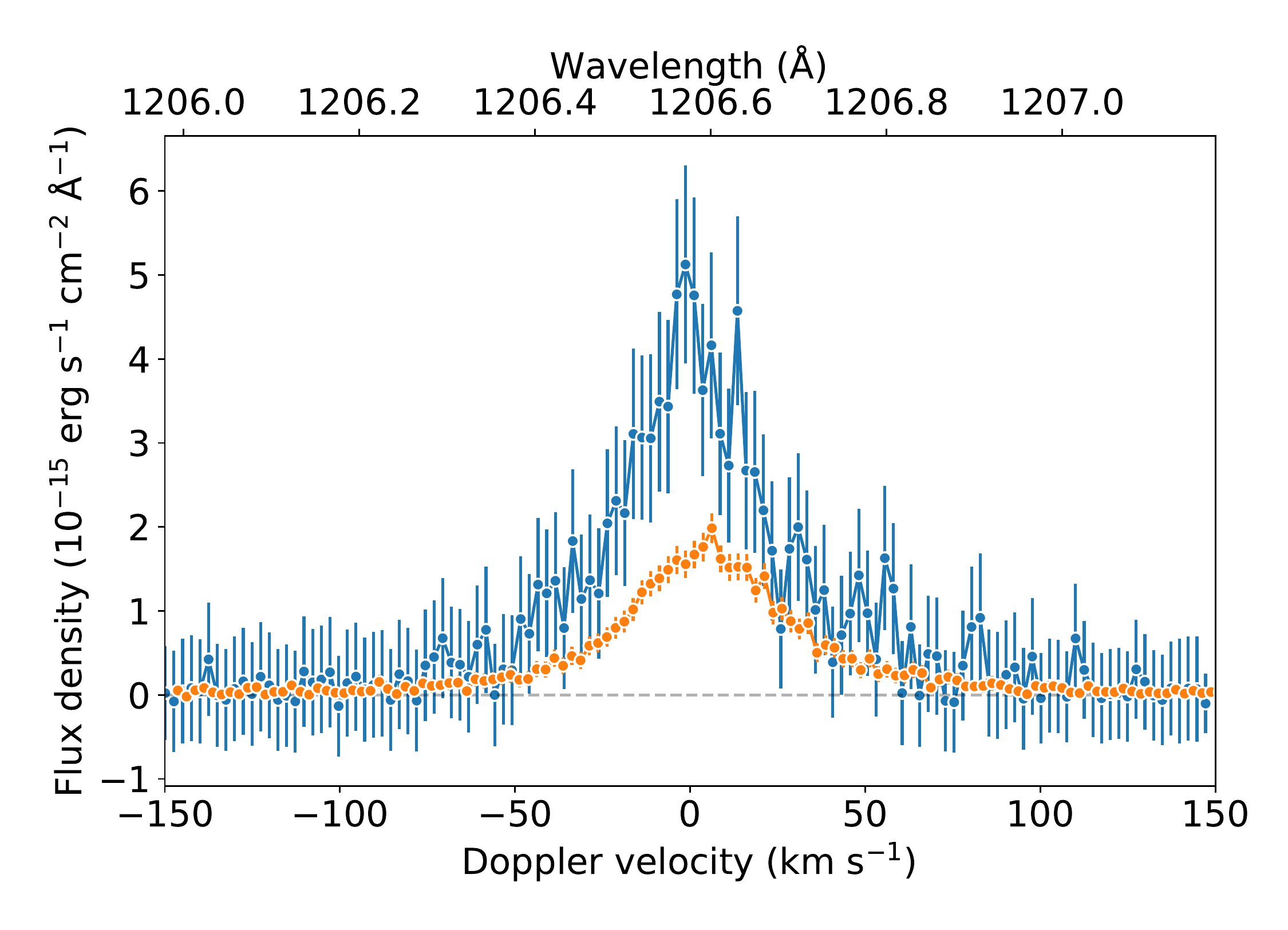} & \includegraphics[width=0.47\hsize]{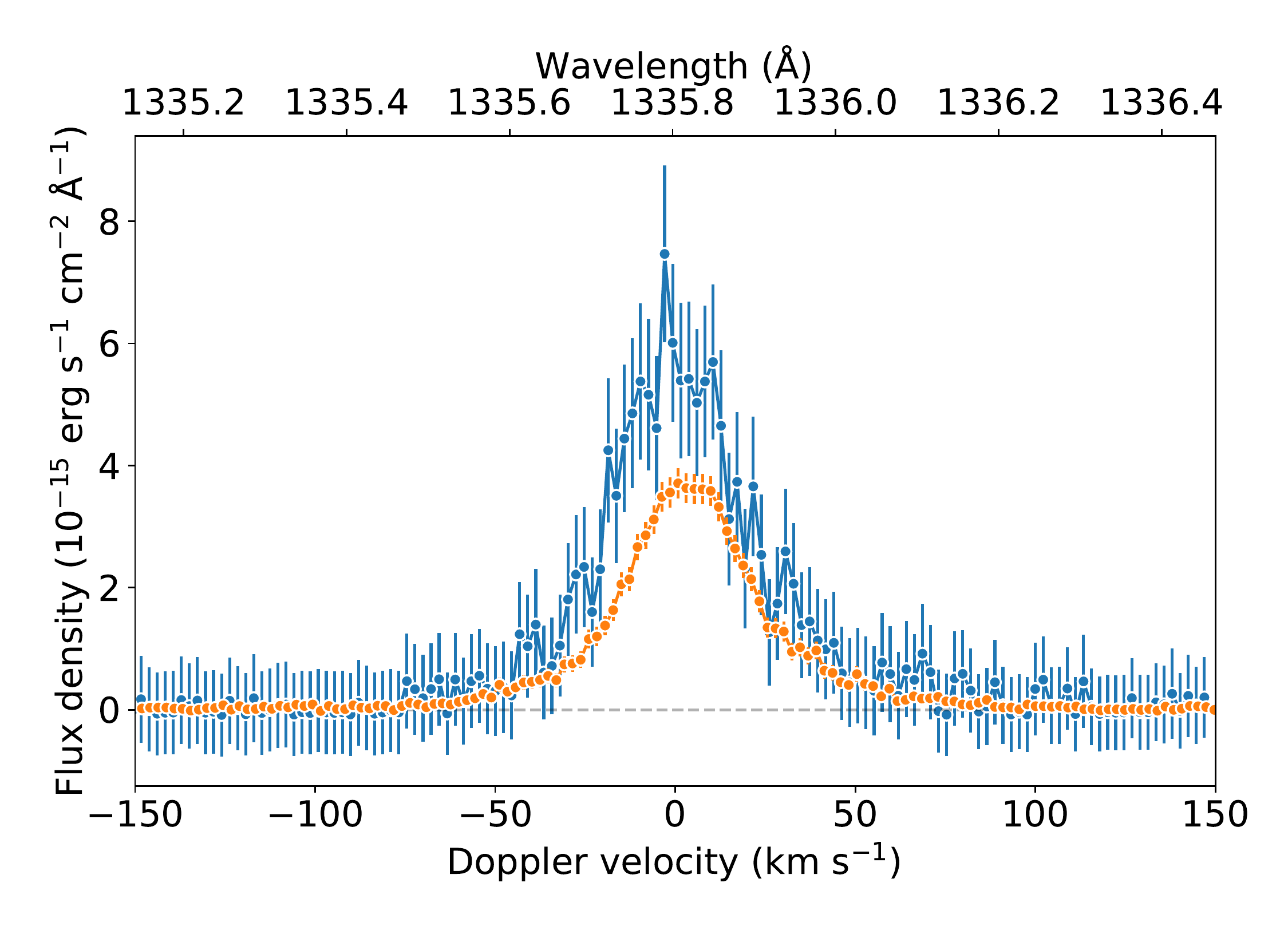} 
        \end{tabular}
    \caption{Comparison between flare spectrum (blue) of GJ~436 against its quiescent spectrum (orange) near the \ion{Si}{III} (left panel) and \ion{C}{II} (right panel) emission lines. These spectra are computed by combining several \emph{HST}-COS exposures. Absolute velocities are in the stellar rest frame but may be affected by biases resulting from the instrument and post-processing; however, the relative Doppler shift between the flare and quiescent spectra is physical.}
    \label{flare_spectra}
    \end{figure*}

A comparison between the quiescent and flare spectra of GJ~436 is shown in Fig. \ref{flare_spectra} and Table \ref{flare_flux}; these spectra were derived by combining all of the out-of-transit, time-tag split subexposures in the quiescent and flare state. The flare spectrum seems to be blueshifted in relation to the quiescent spectrum for both spectral lines shown, which is unexpected given that other M dwarfs and the Sun exhibit a redshifted flare excess instead \citep{2003ApJ...597..535H, 2018ApJ...867...71L}. We presume this blueshift is physical, since we applied wavelength shift corrections for systematic errors uniformly across visits before combining the flare and quiescent spectra; these correction factors are estimated using exposures from which the flare subexposure was eliminated. Stellar lines can show physical redshifts (or even slight blueshifts) because of the chromospheric structure \citep[see, e.g.,][]{2012ApJ...754...69L, 2018A&A...615A.117B}. We cannot measure the absolute position of a given line relatively to the stellar photosphere, but we can measure relative shifts of a flaring line relatively to its quiescent state. The fact that these transition region lines exhibit slightly blueshifted excess is indicative of material flowing upward from the stellar surface.

\begin{table}
\caption{Quiescent and flare fluxes of GJ~436 in lines most sensitive to stellar activity.}
\label{flare_flux}
\centering
\begin{tabular}{l c c c}
\hline\hline
\multirow{2}{*}{Ion} & Central wavelength & Quiescent flux & Flare flux \\
 & ($\mathrm{\AA}$) & \multicolumn{2}{c}{($\times 10^{-16}$ erg s$^{-1}$ cm$^{-2}$)} \\
\hline
\ion{C}{II} & 1334.532 & $2.99 \pm 0.10$ & $4.91 \pm 0.72$ \\
\ion{C}{II} & 1335.708 & $7.99 \pm 0.13$ & $13.5 \pm 0.8$ \\
\ion{Si}{III} & 1206.5 & $4.33 \pm 0.10$ & $10.2 \pm 0.7$ \\
\hline
\end{tabular}
\end{table}

In order to avoid contamination by flares in our results, we removed the subexposures affected by flares from the analysis. The \ion{Si}{III} and \ion{C}{II} lines are the ones that more clearly show these features. In the case of GJ~436, we deem subexposures to be contaminated with flares when the combined fluxes of the \ion{Si}{III} line and the \ion{C}{II} doublet exceed $2.4 \times 10^{-15}$~erg~s$^{-1}$~cm$^{-2}$. Discerning the quiescent flux level from flare-contaminated subexposures from faint emission lines by eye or by analyzing light curves is neither straightforward or reliable. However, we were able to take advantage of a large number of measurements to average out stochastic variations from the baseline flux.

\section{Rotational modulation of FUV fluxes}\label{rot_modulation_sect}

\citet{2018Natur.553..477B} and \citet{2018AJ....155...66L} used photometric observations acquired from 2003 to 2017 with the T12 0.80~m automatic photoelectric telescope (APT) at the Fairborn Observatory \citep{1999PASP..111..845H}. They found evidence for rotational modulation with a period of 44.09 days and a peak-to-peak amplitude of 0.0032 mag. In this section we quantify the potential modulation of the stellar FUV fluxes due to the presence of active regions on the stellar surface of GJ~436.

\subsection{Assessing the presence of a long-lived active region or an active longitude}

In order to verify the presence of rotational modulation of the fluxes in the FUV lines of GJ~436, we phase-folded the data to the rotational period of the star and fit the fluxes with the following two different models: a sinusoid and a constant. Further, we used the Bayesian Information Criterion \citep[BIC;][]{schwarz197801} to evaluate which of these two models best describe the overall behavior of the fluxes when phase-folded to the rotational period of the star.

We fit the flux modulation by maximizing the likelihood function

\begin{equation}
\ln \left[p\left(F_\mathrm{obs} | \mathcal{M} \right) \right] = - \frac{1}{2} \sum_{k} \left[ \frac{\left(F_{\mathrm{obs}, k} - F_{\mathcal{M}, k}\right)^2}{\sigma_k^2} + \ln \left(2 \pi \sigma_k^2 \right)\right] \mathrm{,}
\end{equation}where $F$ is the flux, $\mathcal{M}$ is the model, and $\sigma$ is the uncertainty of the flux. The best fit was calculated using the truncated Newton algorithm implementation of SciPy \citep{scipy_ref}. The fit parameters are the amplitude, reference phase, and flux baseline for the sinusoidal model; the flux baseline was the only parameter used for the constant model (the amplitude and baseline were measured as a fraction of the mean observed flux). The rotational period was fixed at 44.09 d. We did not use in-transit fluxes in this analysis to avoid contamination by possible transit signals. We performed the fits for the combined flux of multiplets where applicable in order to have the highest signal-to-noise ratio possible for each species. We evaluated the hypothesis of sinusoidal modulation using the Bayesian Information Criterion (BIC).

M dwarfs can display long-lived active regions that modulate their fluxes over several rotations. It is not clear, however, how long-lived they are on GJ~436; the extended data in Fig. 7 from \citet{2018Natur.553..477B} show that the low-amplitude rotational modulation of GJ~436 stays coherent over 14 consecutive years in optical wavelengths. On visual inspection, the rotationally phase-folded fluxes of \ion{Si}{III}, \ion{C}{II,} and \ion{N}{V} seem to modulate with a sinusoidal behavior, except for measurements during the mid-2012 epoch. Although the measurements from the mid-2015 epoch follow the trend, it is also plausible that they are coincidentally higher near a magnetic cycle maximum and are not necessarily related to an active region being observed in the late-2017 to early-2018 epoch. We thus propose the following two hypotheses to be assessed: (a) the measurements during epochs mid-2012 and mid-2015 are not coherent with the latest epoch and should not be included in the rotational modulation analysis; (b) the mid-2015 epoch is coherent with the latest epoch, either by an active longitude \citep[as seen in the Sun and in GJ~1214;][]{2003A&A...405.1121B, 2005AstL...31..280K, 2013ApJ...770..149W, 2018A&A...614A..35M} or the same long-lived active region, and it should be included in the rotational modulation analysis.

Assuming hypothesis (b), we found that the fluxes of the \ion{C}{II} doublet, the \ion{N}{V} doublet, and the \ion{Si}{III} line in the COS data seem to display rotational modulation with a sinusoidal model favored by $\Delta \mathrm{BIC} > 10$ in relation to the constant model (see Fig. \ref{rot_modulation}). The flux time series of the individual species with weaker lines are not as well described by sinusoids but, when their fluxes are combined, the modulation is clear (lower right panel of Fig. \ref{rot_modulation}). We used a Markov-Chain Monte Carlo (MCMC) algorithm to evaluate the uncertainties of the sinusoidal fit and verified that it displays reference phases that are similar within 1$\sigma$ given the uncertainties of the fit. 

\begin{figure}
\centering
\includegraphics[width=\hsize]{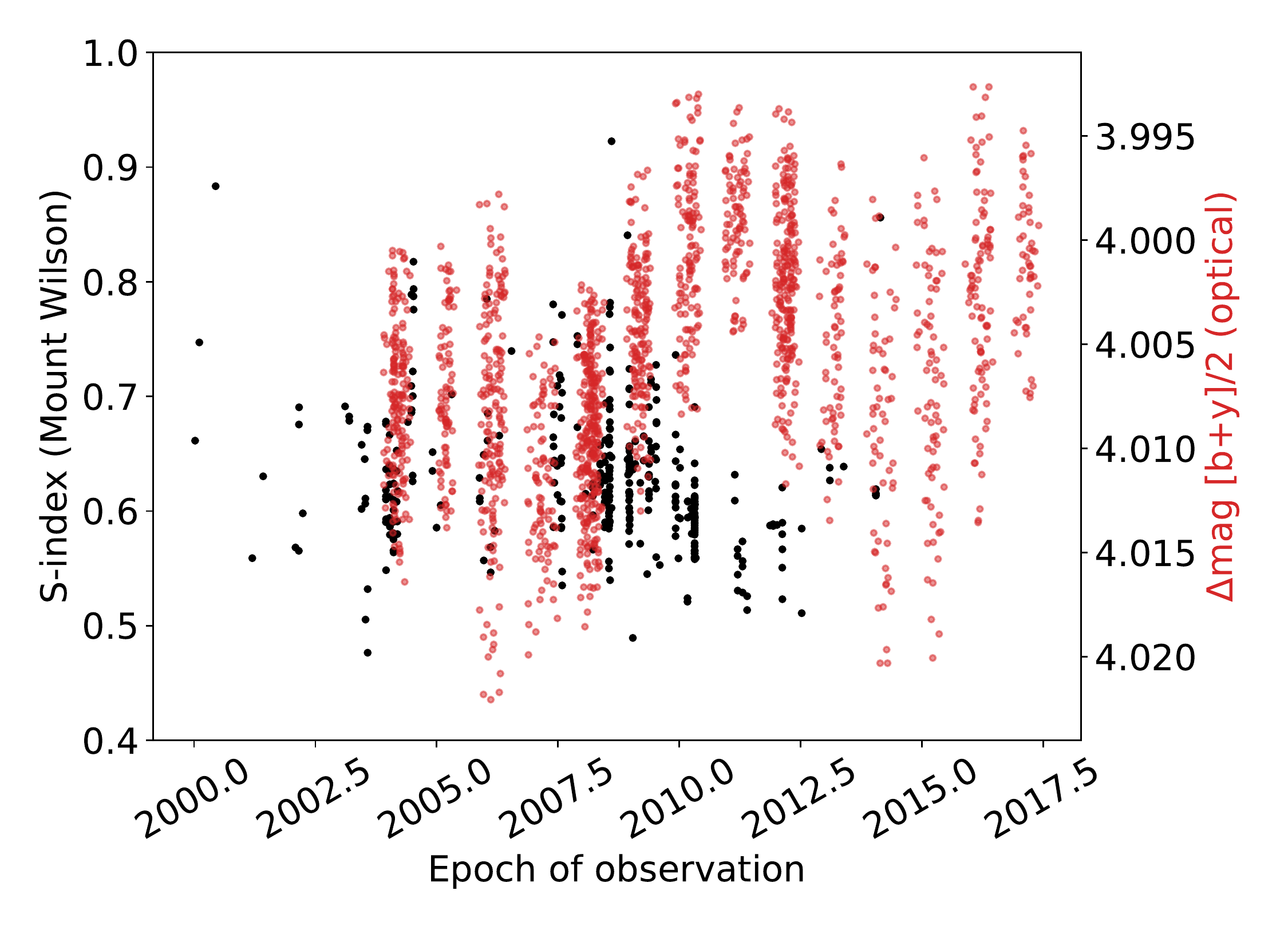}
\caption{Long-term activity modulation of GJ~436 seen in optical photometry (red) and $S$-index (black).}
\label{GJ436_phot}
\end{figure}

\begin{figure*}[ht]
\centering
\begin{tabular}{cc}
\includegraphics[width=0.47\hsize]{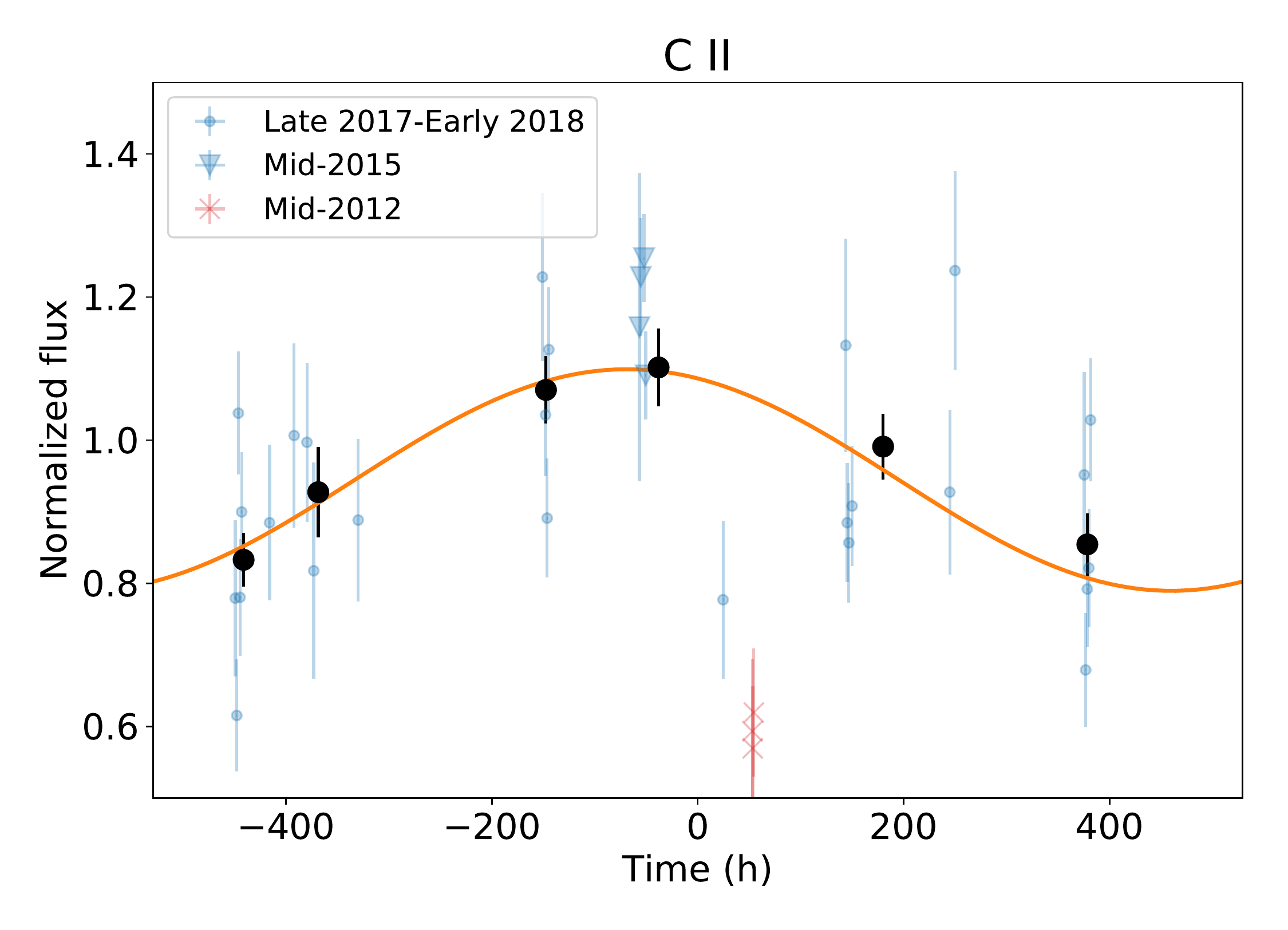} & \includegraphics[width=0.47\hsize]{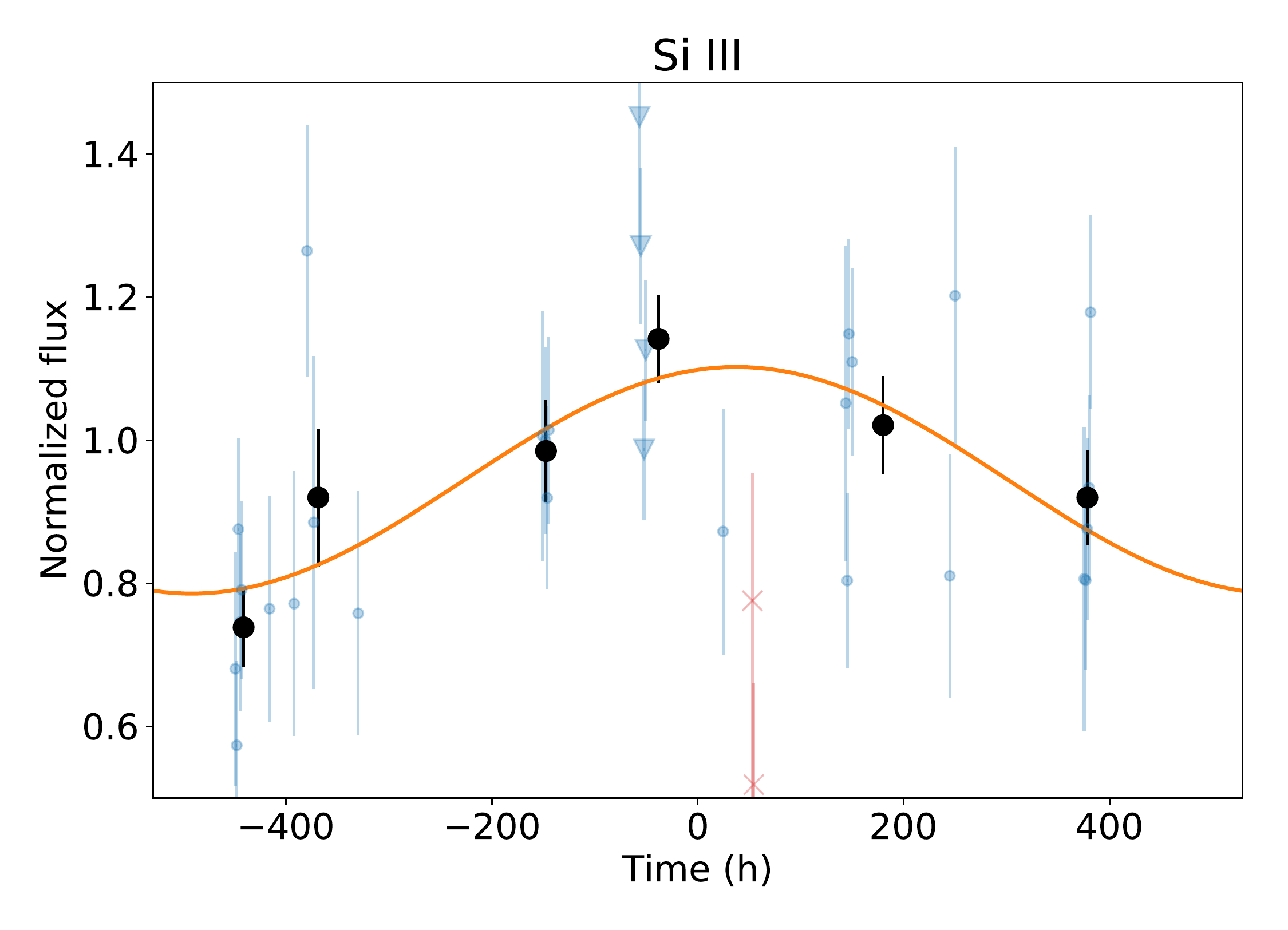} \\
\includegraphics[width=0.47\hsize]{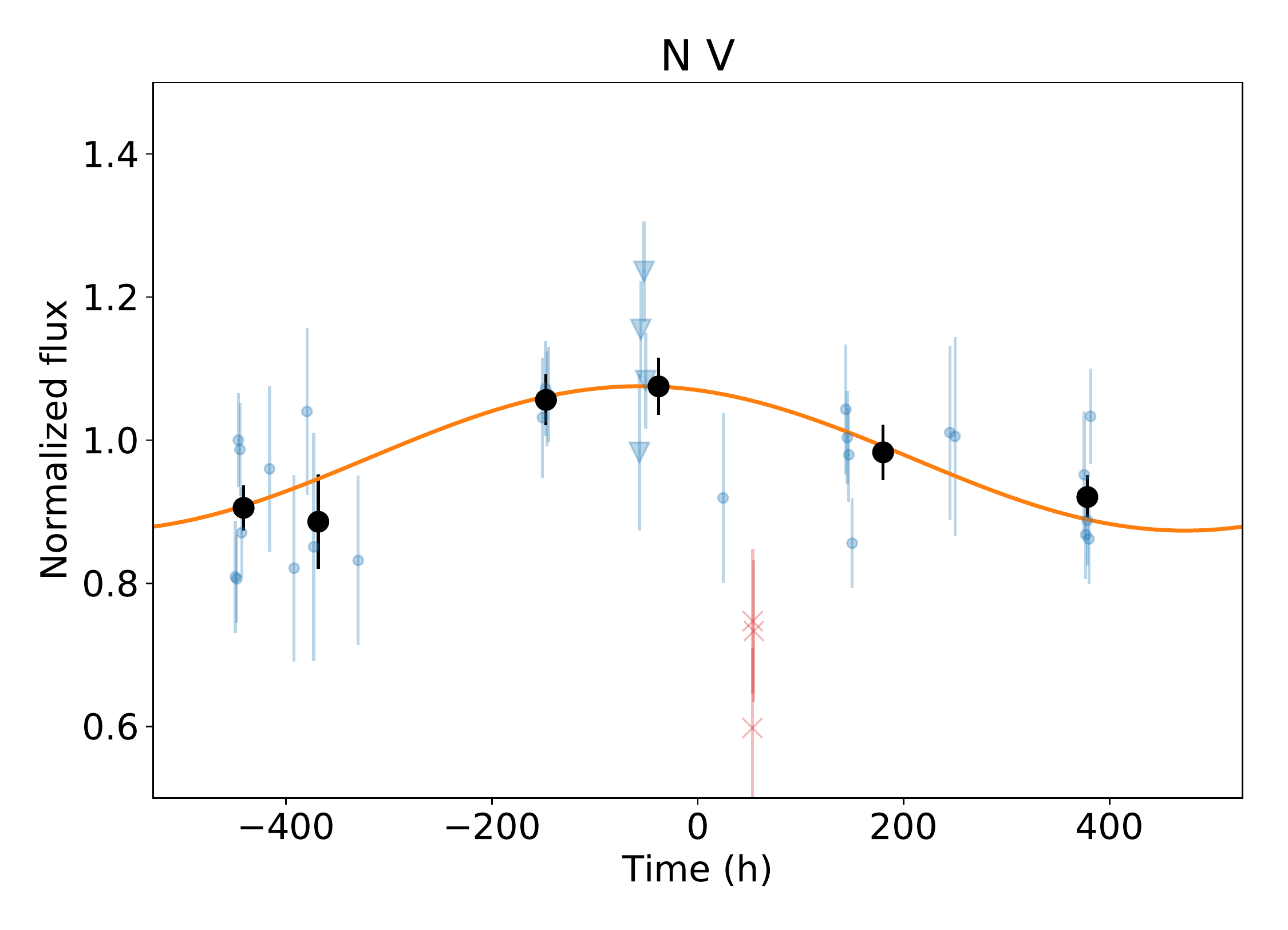} & \includegraphics[width=0.47\hsize]{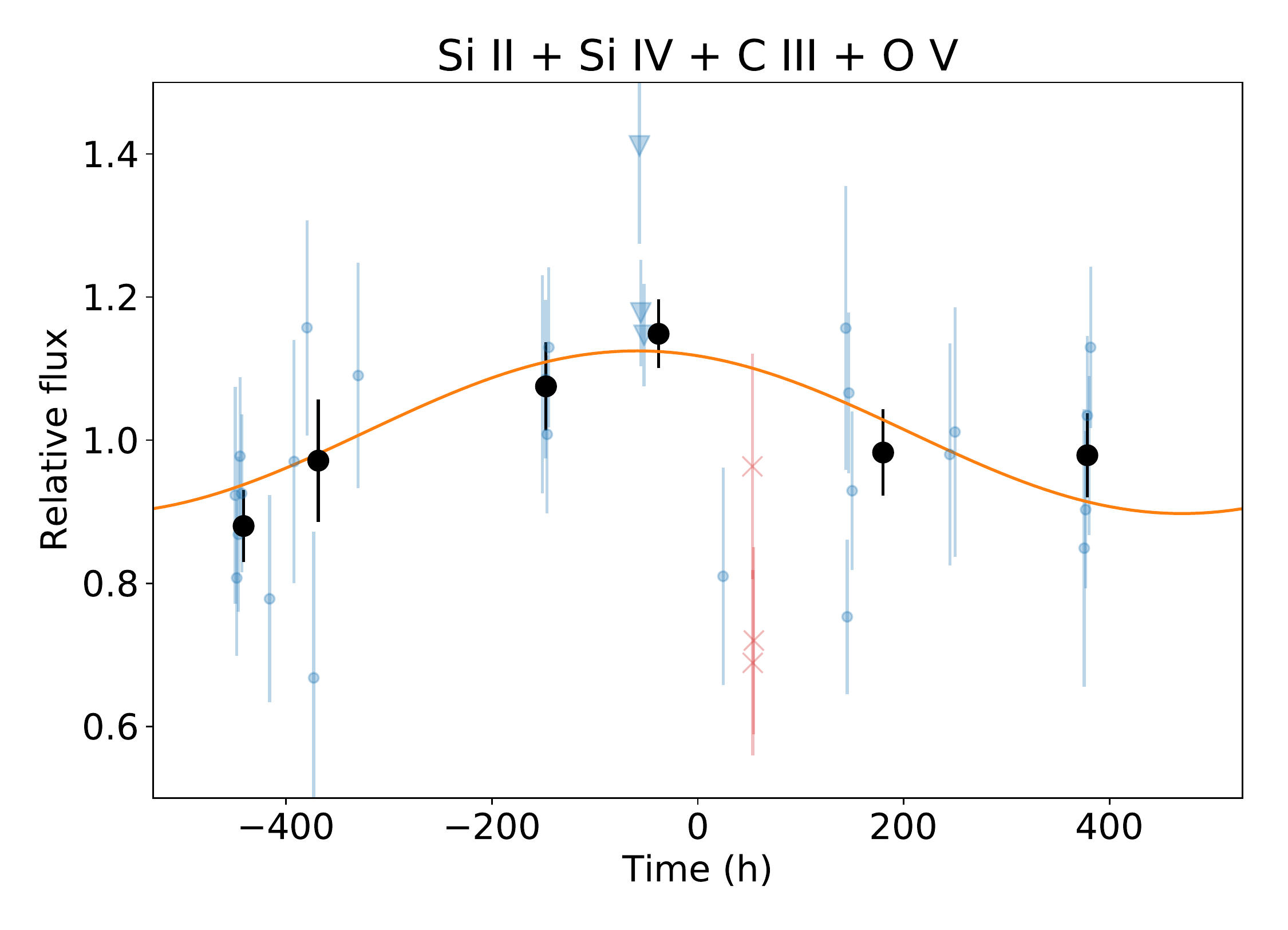} \\
\end{tabular}
\caption{Rotational modulation of fluxes of \ion{C}{II} doublet, \ion{Si}{III} line, \ion{N}{V} doublet, and combined fluxes of \ion{Si}{II}, \ion{Si}{IV}, \ion{C}{III,} and \ion{O}{V} lines in the COS spectra, assuming that epoch mid-2015 is affected by the same active region or longitude. The light curves were phase-folded to the rotational period of the star GJ~436 (44.09 d); the amplitude and baseline are measured in fraction of the mean average flux. The black data points are bins of groups of observations near the same phase. The fits were performed to the orbit-to-orbit data and not to the binned data.}
\label{rot_modulation}
\end{figure*}

Assuming hypothesis (a), we found that the sinusoidal rotational modulation model is not significantly favored over a constant flux model ($\Delta \mathrm{BIC} < 10$). In this case, it would not be necessary to correct for rotational modulation during the light curve analysis. On the other hand, the epochs from mid-2012 and mid-2015 cannot be included in the light curve analysis because we are not able to accurately assess the effects of activity in the flux measurements during these earlier epochs. A visual inspection of the $S$-index of activity \citep{1978PASP...90..267V} of GJ~436 measured with the HIRES spectrograph \citep{2017AJ....153..208B} suggests that the star was at a minimum of its activity cycle around 2012, and that the activity started to increase again around 2014. A consistent behavior is also seen in optical photometric monitoring of GJ~436; based on fig. 1 of \citet{2018AJ....155...66L}, the epoch when the star becomes the brightest in optical, which represents the minimum spot coverage, roughly corresponds to the minimum of $S$-index around 2012 (see Fig. \ref{GJ436_phot}). Furthermore, the epoch of lowest optical flux (between 2014 and 2016), representing maximum spot coverage, corresponds to the epochs when we see the highest UV fluxes in the \emph{HST} data. The stellar magnetic cycle modulation\footnote{\footnotesize{\citet{2018AJ....155...66L} found that the magnetic cycle of GJ~436 is roughly 7.4 years.}} can explain the variation seen between the observations in epochs in mid-2012 and mid-2015.

As the strongest emission line in the FUV spectra of M dwarfs, the Lyman-$\alpha$ line is also susceptible to rotational modulation in its flux. In the case of GJ~436, the blue wing of the line (Doppler velocities range [-120, +50] km s$^{-1}$) is affected by a strong and long-lasting planetary absorption when near or in-transit. However, the atmospheric escape models of \citet{2016A&A...591A.121B} suggest that the far blue wing ([-250, -120] km s$^{-1}$) and the reference red wing ([+120, +250] km s$^{-1}$) of the Lyman-$\alpha$ line of GJ~436 should be free of planetary signals (see Sect. \ref{lya_cos}).

\begin{figure}[h]
    \centering
    \includegraphics[width=\hsize]{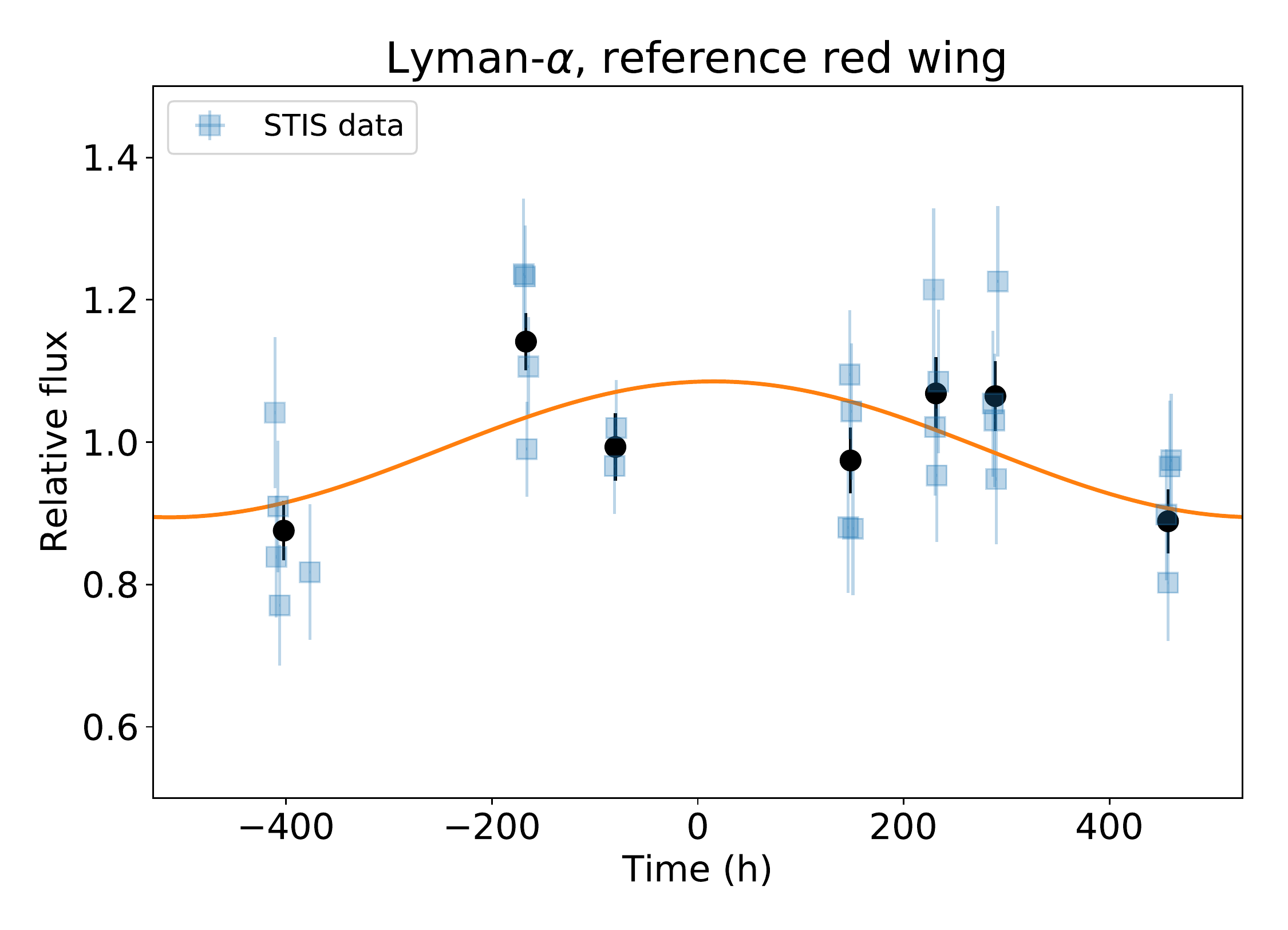}
    \caption{Same as Fig. \ref{rot_modulation}, but for fluxes in Lyman-$\alpha$ reference red wing [120, 250] km s$^{-1}$ as observed with STIS spectrograph. The epoch of observations span from 2010 to 2016.}
    \label{rot_modulation_lya}
\end{figure}

We re-analyzed the STIS data obtained in previous programs \citep{2014ApJ...786..132K, 2015Natur.522..459E, 2017A&A...605L...7L} to check for rotational modulation in the reference red wing of the Lyman-$\alpha$ line of GJ~436 (Doppler velocities range [120, 250] km s$^{-1}$). The epochs of observation with the STIS spectrograph range from mid-2010 to early-2016, which encompasses the supposed activity minimum around 2012 and the increase in activity starting in 2014. In this case, we found that the sinusoidal fit is not significantly favored against the constant flux model ($\Delta \mathrm{BIC} \approx 7$; see Fig. \ref{rot_modulation_lya}).

\begin{table}
    \caption{Bayesian information criterion (BIC) and deviance ($\chi^2$) values of models for stellar flux variation.}
    \label{bic_chisq}
    \centering
    \begin{tabular}{l c c c c c}
    \hline\hline
    \multirow{2}{*}{Species} & \multirow{2}{*}{Epochs} & \multicolumn{2}{c}{Sinusoidal} & \multicolumn{2}{c}{Constant} \\
     & & BIC & $\chi^2$ & BIC & $\chi^2$ \\
    \hline
    \multirow{2}{*}{\ion{C}{II}} & 2015-2018 & -65.5 & 63.3 & -25.2 & 110 \\
    & 2017-2018 & -56.9 & 49.1 & -51.8 & 60.8 \\
    \hline
    \multirow{2}{*}{\ion{Si}{III}} & 2015-2018 & -69.8 & 35.1 & -54.8 & 57.0 \\
    & 2017-2018 & -58.5 & 25.4 & -55.8 & 34.5 \\
    \hline
    \multirow{2}{*}{\ion{N}{V}} & 2015-2018 & -107 & 32.6 & -84.9 & 61.8 \\
    & 2017-2018 & -93.8 & 23.5 & -88.9 & 34.9 \\
    \hline
    All weaker & 2015-2018 & -81.6 & 32.7 & -71.7 & 49.3 \\
    lines & 2017-2018 & -71.1 & 22.0 & -76.4 & 23.2 \\
    \hline
    Lyman-$\alpha$ & 2010-2016 & -78.3 & 42.1 & -71.5 & 55.4 \\
    \hline
    \end{tabular}
    \end{table}

\subsection{Discussion}\label{rot_discussion}

To the best of our knowledge, the potential modulation of FUV fluxes that we found for GJ~436 has not been previously reported for another quiet M dwarf in the literature. Using optical photometric data obtained with the \emph{Kepler} mission, \citet{2017MNRAS.472.1618G} concluded that M dwarfs with rotational periods of 10 or 20 d can have active regions with lifetimes varying from a few tens of to 430 days, depending on the size of the active region (in general, larger regions should last longer than smaller ones). Furthermore, \citet{2015ApJ...801...79R} reported on the presence of a large active region or complex of spots in GJ~176, an M2 dwarf with a rotational period of 39 d, and this region remained stable for at least six years (the photometric variability remained in phase during the span of observations, unlike the other activity indices). Another example is Proxima Cen, which possesses a longer rotational period and similar activity index as GJ~436 ($P_\mathrm{rot} = 83.2$ d and $\log{R'_\mathrm{HK}} = -5.65$) but exhibits rotational modulation that is stable for more than eight years \citep{2015MNRAS.452.2745S, 2016A&A...595A..12S}.

If hypothesis (b) is correct, then our results indicate that GJ~436 possessed either a stable active region or an active longitude that modulated the FUV fluxes for more than 45 rotations since the mid-2015 epoch. In principle, we do not expect the rotational modulation in FUV spectra to be in phase with optical broadband photometry since they trace different regions of the stellar atmosphere. In fact, we expect them to be out of phase by $\pi / 2$ since active regions usually display bright features in the ultraviolet \citep[e.g.,][]{1973ApJ...182..321D, 2000ApJ...543.1016B} and dark spots in broadband optical wavelengths \citep[e.g.,][]{1671RSPT....6.2295H, 1997MNRAS.287..556C, 2018MNRAS.474.5534O}.

If hypothesis (a) is correct, then the flux modulations seen in epochs in mid-2012 and mid-2015 are not related to a putative active region observed in the later epochs (late-2017 to early-2018), and they are more likely related to the longer magnetic cycle of the star instead of rotation. In addition, the observations from the later epochs do not display significant rotational modulation by themselves. It is difficult to disentangle hypotheses (a) and (b) because similar observations have not been performed between mid-2015 and late-2017, and both of them have observational evidences in their favor. \citet{2016ApJ...824..101Y} also investigated the variability of the Lyman-$\alpha$ of several M dwarfs and found no significant variability for GJ~436; other, more active M dwarfs did exhibit variations in their Lyman-$\alpha$ flux in the order of 10-20\%.

\section{The planetary Lyman-$\alpha$ absorption}\label{lya_cos}

Previous observations with \emph{HST}-STIS showed that GJ~436~b possesses an extended H-rich exosphere, which is readily detectable in Lyman-$\alpha$ transit spectroscopy \citep{2014ApJ...786..132K, 2015Natur.522..459E, 2015A&A...582A..65B, 2016A&A...591A.121B, 2017A&A...605L...7L}. In this section we analyze the Lyman-$\alpha$ time-series of GJ~436~b obtained during several transits with \emph{HST}-COS in order to reproduce this previous detection and verify the suitability of geocoronal emission correction for faint FUV targets.

\subsection{Airglow contamination correction}\label{ag_corr}

The FUV spectra obtained with \emph{HST} are contaminated by geocoronal emission (also known as airglow) in the Lyman-$\alpha$ (1215.6702 \AA) and \ion{O}{I} lines (1302.168, 1304.858, and 1306.029 \AA). The level of contamination is variable and tends to either increase (decrease) during an orbit depending on whether the telescope is moving from the Earth's night(day)-side to day(night)-side. This variation is visible when the time-tag data are split in subsequent subexposures. The level of contamination also varies from orbit to orbit and, in our data, the first orbit of a given visit tends to be more severely contaminated; this is because the first orbit usually started closer to the Earth's dayside than the other orbits.

When using STIS, the instrument pipeline automatically removes the geocoronal contamination by taking advantage of the fact that it is a slit spectrograph; in the case of COS, which has a circular aperture, it is impossible to measure the airglow independently from the stellar spectra, so a simple automatic subtraction of contamination is not possible. However, \citet{2018A&A...615A.117B} showed that it is possible to correct the Lyman-$\alpha$ emission of 55 Cnc on \emph{HST}-COS spectra and remove the geocoronal contamination by using airglow templates\footnote{\footnotesize{Airglow templates for \emph{HST}-COS are available at \url{http://www.stsci.edu/hst/cos/calibration/airglow.html}.}} accumulated from previous programs \citep[see also][]{2013A&A...553A..52B, 2017A&A...599A..75W}. Here, we applied the same technique to the GJ~436 spectra. Since the \ion{O}{I} lines of GJ~436 are too faint to be discerned from the airglow emission, we decided to perform the correction only for the Lyman-$\alpha$ line and discard the \ion{O}{I} lines analysis.

\begin{figure}
\includegraphics[width=1.00\hsize]{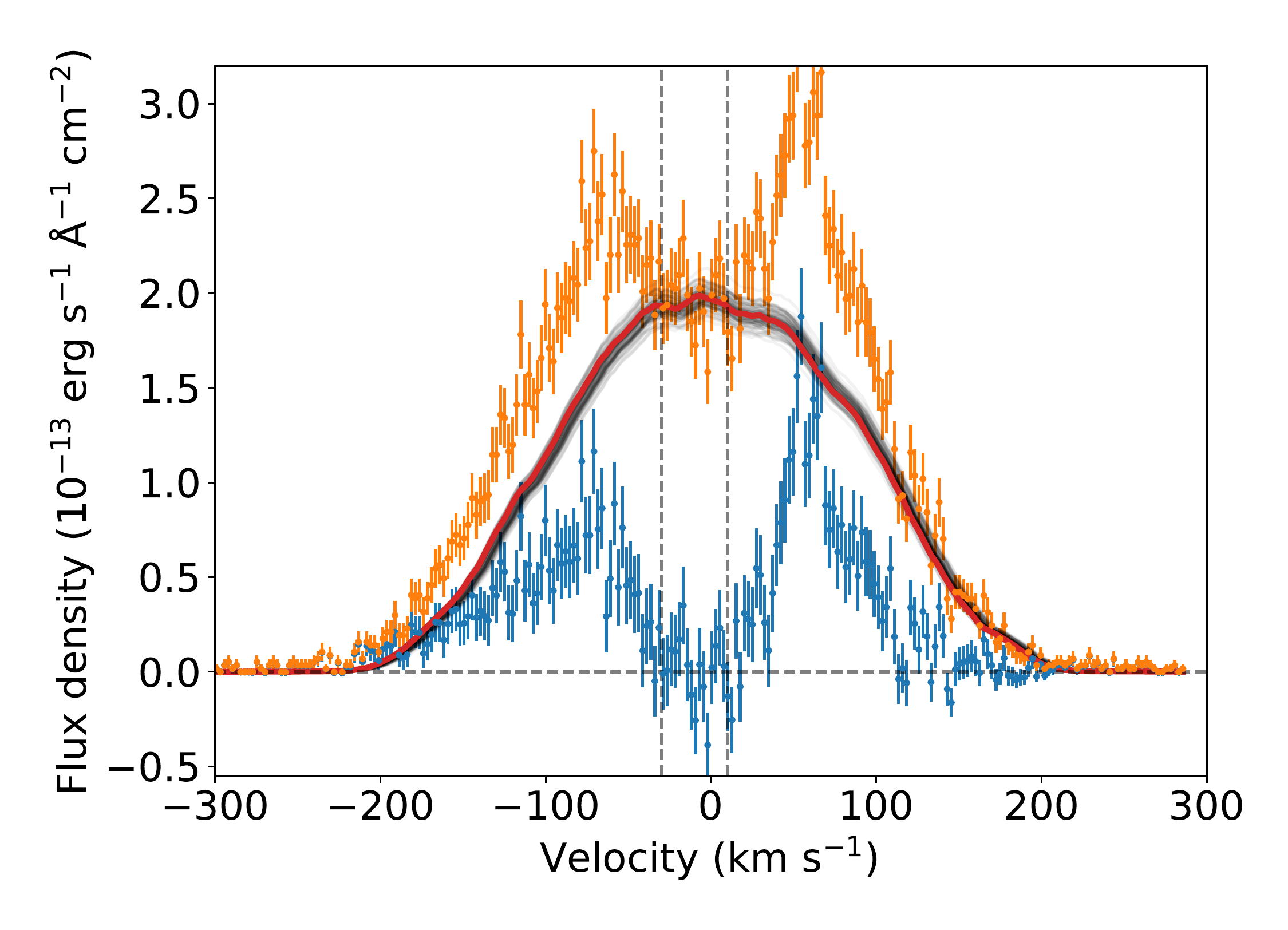}
\caption{Airglow contamination removal in Lyman-$\alpha$ profile of GJ~436 measured during orbit 1 of Visit D. The cleaned (contaminated) spectrum is shown in blue (orange), the best fit airglow template is shown in red, and the MCMC posterior sample is shown as a family of gray airglow templates.}
\label{airglow}
\end{figure}

The geocoronal airglow spectra have an approximately constant shape. In order to subtract the airglow from the observations, we needed to fit the amplitude and Doppler shift of the template to the observed emission line since the observed spectra have systematic Doppler shifts that we needed to correct for. Following the procedure outlined in \citet{2018A&A...615A.117B}, we fit the core of the airglow template to a region of the observed Lyman-$\alpha$ profile where we did not expect any emission from the star. The wavelength range where the interstellar medium (ISM) completely absorbs the stellar emission line\footnote{\footnotesize{See estimates for the \ion{H}{I} column density in the line of sight of GJ~436 in \citet{2015A&A...582A..65B}.}} is suitable to fit the amplitude of the airglow; in our data set, this spectral range is located between -10 and 30 km s$^{-1}$ in the rest frame of the star \citep{2015A&A...582A..65B}. The best fit is obtained by minimizing an objective function, namely the difference between the observed spectrum and the template in the aforementioned range using a truncated Newton algorithm. The objective function also includes a term that penalizes airglow templates that produce negative fluxes when subtracted from the observed spectra. The fit parameters are the Doppler shift of the airglow in relation to the stellar spectra and the amplitude of the airglow in each exposure. We estimated the uncertainties of the fit by performing a MCMC simulation; an example of the airglow removal results is shown in Fig. \ref{airglow}.

Since the data were obtained in time-tag mode, each exposure was divided in four in order to select the subexposures with the least geocoronal contamination (i.e., near the Earth's night side). As long as the wings of the stellar Lyman-$\alpha$ emission can be visually distinguished from the geocoronal emission, then the subexposure is suitable for airglow removal. The datasets that could be corrected are listed in Table \ref{airglow_log}.

\begin{table}
\caption{Summary of suitability of COS subexposures for Lyman-$\alpha$ recovery. Each subexposure corresponds to a quarter of the total exposure in the orbit. Subexposures marked with \checkmark are suitable for Lyman-$\alpha$ recovery; those marked with $\times$ are contaminated by flares.}
\label{airglow_log}
\centering
\begin{tabular}{l c | c | c | c | c}
\hline\hline
Visit & Orbit & \multicolumn{4}{c}{Subexposures} \\
\hline
\multirow{4}{*}{A} & 1 & & & & \\
& 3 & \checkmark & & & \\
& 4 & \checkmark & & & \\
& 5 & & & & \checkmark \\
\hline
\multirow{5}{*}{B} & 1  & & & & \\
& 2 & \checkmark & & & \\
& 3 & & & & \checkmark \\
& 4 & \checkmark & & & \\
& 5 & \checkmark & & & \\
\hline
\multirow{5}{*}{C} & 1 & $\times$ & \checkmark & & \\
& 2 & & \checkmark & \checkmark & \checkmark \\
& 3 & & \checkmark & \checkmark & \checkmark \\
& 4 & & \checkmark & \checkmark & \checkmark \\
& 5 & \checkmark & \checkmark & \checkmark & \\
\hline
\multirow{4}{*}{D} & 1 & & $\times$ & & \checkmark \\
& 2 & \checkmark & & \checkmark & \checkmark \\
& 3 & & \checkmark & \checkmark & \checkmark \\
& 5 & \checkmark & & \checkmark & \checkmark \\
\hline
\end{tabular}
\end{table}

\begin{figure}
\includegraphics[width=1.00\hsize]{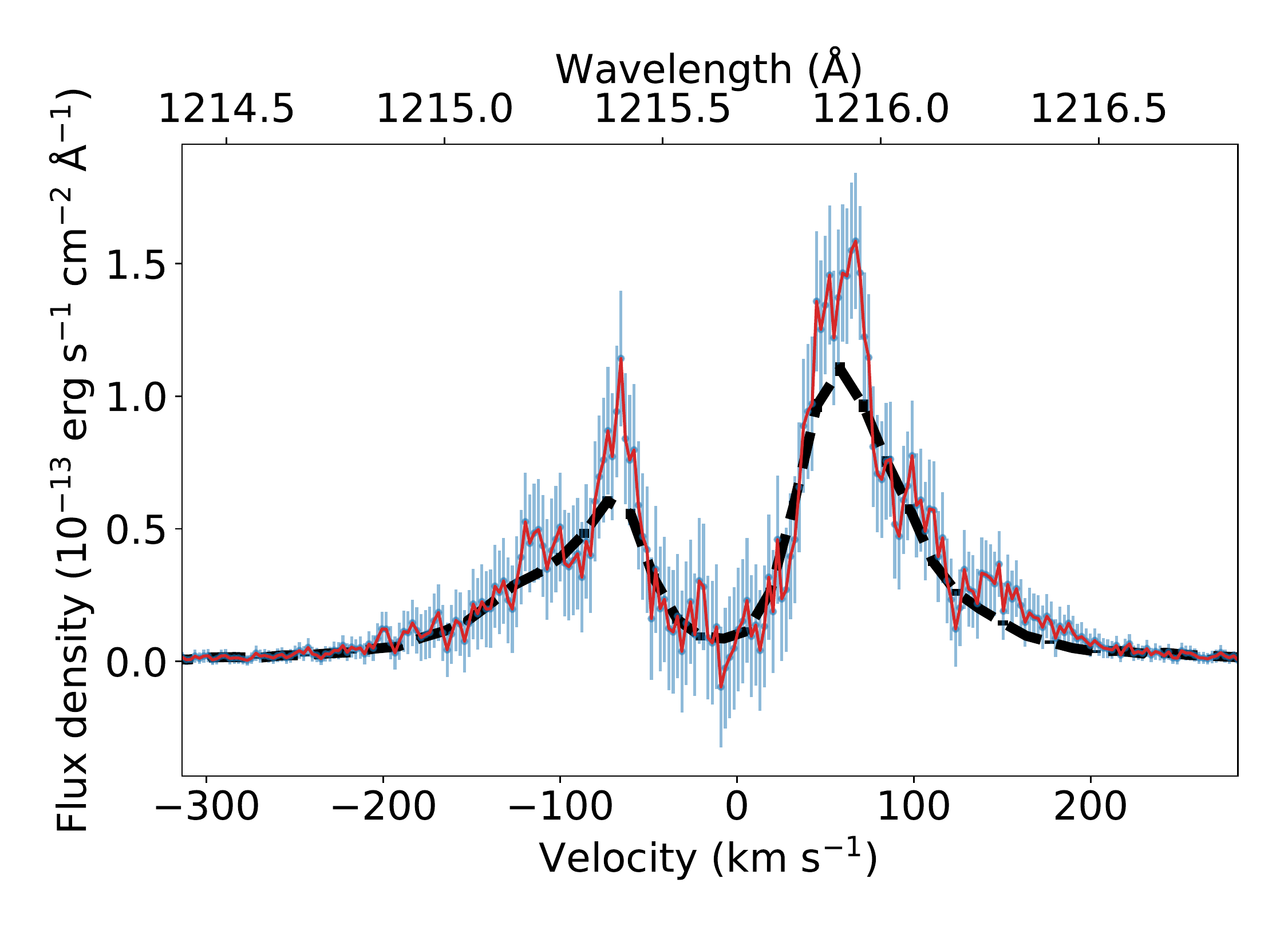}
\caption{Mean out-of-transit Lyman-$\alpha$ spectrum of GJ~436 measured with \emph{HST}-COS observations from program GO-15174 (red spectrum with blue uncertainty bars). For comparison, we plot the mean out-of-transit spectrum measured with \emph{HST}-STIS in black. The spectra are centered in the stellar rest frame. The intrinsic Lyman-$\alpha$ emission line of GJ~436 likely possesses a single-peaked Voigt shape \citep{2015A&A...582A..65B}, but the core of the line is absorbed by the ISM, producing a double-peaked feature when observed from the Earth.}
\label{lya_oot}
\end{figure}

We measured the out-of-transit spectrum of GJ~436 with \emph{HST}-COS observations from orbits 5 and 7 of program GO-15174 (see Fig. \ref{lya_oot}; the other orbits were either too close to the planetary transit or too contaminated by airglow). The COS exposures of the MUSCLES dataset were not used in the Lyman-$\alpha$ analysis because part of the airglow profile falls inside the region with shadows caused by the wire grid of COS (the shadows produce a 15\% depression in the continuum). The pipeline normally calibrates these regions as point sources during the flat-field correction; however, the Earth's airglow is not a point source, so this correction is not perfect if the emission falls in this region, resulting in spurious emissions on top of the airglow.

\subsection{Stable absorption in the blue wing}\label{blue_wing}

\begin{figure}
\includegraphics[width=1.00\hsize]{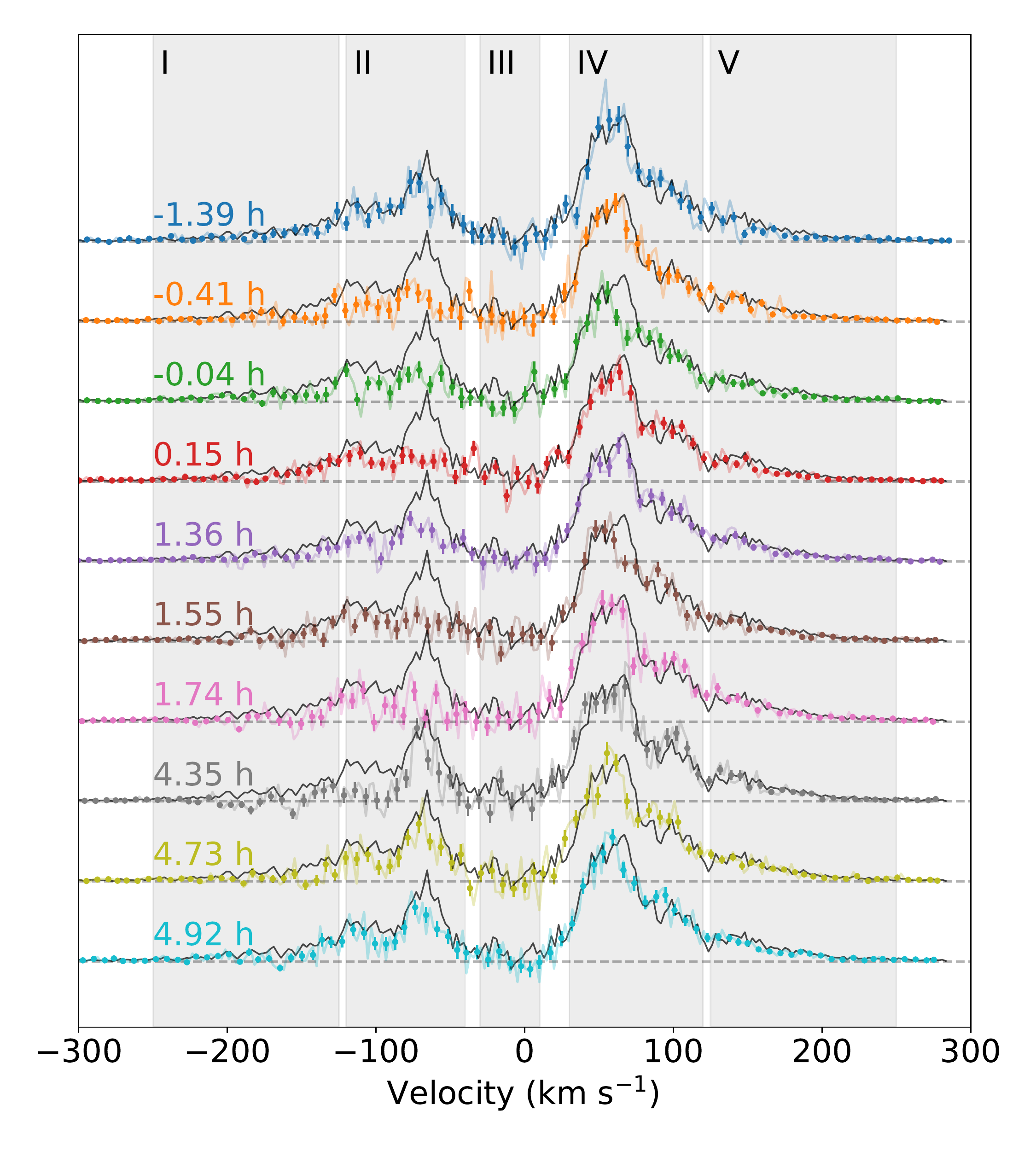}
\caption{\emph{HST}-COS Lyman-$\alpha$ profiles of GJ~436 observed during Visit D after airglow decontamination (color spectra), binned to Doppler velocity intervals of 7 km s$^{-1}$. The reference out-of-transit spectrum is plotted for comparison as a black spectrum against each exposure of Visit D. The regions shaded in gray correspond to: I) the reference far blue wing [-250, -120] km s$^{-1}$; II) the blue wing [-120, -40] km s$^{-1}$; III) the line core absorbed by the ISM [-30, +10] km s$^{-1}$; IV) the red wing [+30, +120] km s$^{-1}$; V) the reference far red wing [+120, +250] km s$^{-1}$. We do not expect planetary signals in regions I and V, so they can be used to estimate the stability of the Lyman-$\alpha$ emission. The timestamps correspond to the phases in relation to the orbital motion of the planet.}
\label{lya_pr}
\end{figure}

After applying the geocoronal contamination removal described in Sect. \ref{ag_corr} to the spectra, we obtained the clean Lyman-$\alpha$ profile of GJ 436 during the four visits (see the cleaned spectra from Visit D in Fig. \ref{lya_pr}). More information about the observed and intrinsic shape of the Lyman-$\alpha$ line for a range of stellar types can be found in, \citet{2005ApJS..159..118W}, for example. The variability seen in the line is partly due to photon noise, imperfections in the airglow decontamination, and potential astrophysical signals. The blue wing of the line, inside the Doppler velocity interval [-120, -40] km s$^{-1}$ (region II in Fig. \ref{lya_pr}), is known to display a periodic absorption due to the transit of GJ~436~b and its extended exosphere.

\begin{figure*}
\centering
\includegraphics[width=1.00\hsize]{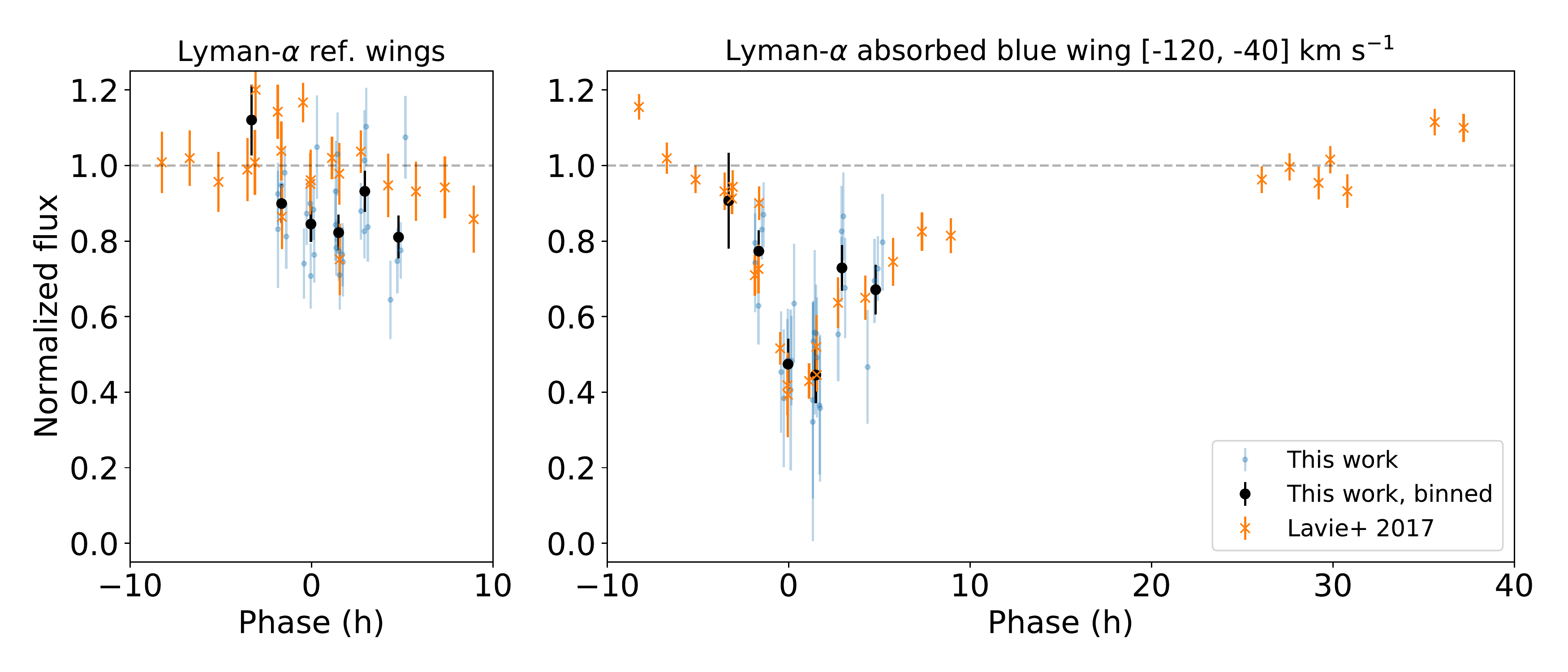}
\caption{Normalized Lyman-$\alpha$ light curves of GJ~436 from COS (this work) and STIS \citep{2017A&A...605L...7L} observations. The baseline fluxes used for normalization were measured in phases outside the [-3, +24] h range. In the case of COS spectra, we measured the baseline flux from the out-of-transit exposures from program GO-15174. The uncertainties of the baseline flux were propagated to the final uncertainties of the normalized fluxes. The left panel shows the light curve of the Lyman-$\alpha$ reference far wing fluxes, measured in the passbands I and V from Fig. \ref{lya_pr}.}
\label{lya_lc}
\end{figure*}

We reproduced the light curve of the Lyman-$\alpha$ blue wing obtained with the COS spectrograph and plotted the previous STIS results in Fig. \ref{lya_lc} for a comparison of the shape of the light curve. Even though we analyzed the same passbands, there may be an offset between COS and STIS passband fluxes due to their different instrumental profiles. The spectral resolving power of STIS/G140M and COS/G130M are, respectively, $\sim$12,000 and $\sim$14,000 near the Lyman-$\alpha$ wavelength. Although the full-line Lyman-$\alpha$ fluxes are expected to be equal, independent of the instrumental profiles, narrower passbands in this line are expected to produce different fluxes between different instruments. In order to avoid these offsets, we show the Lyman-$\alpha$ light curves normalized in relation to the baseline fluxes measured outside the phase range [-3, +24] h.

The planetary absorption in the blue wing of the Lyman-$\alpha$ line of GJ~436 shows that the signal is also present in the COS data, and it remains repeatable over several epochs during our observations (Fig. \ref{lya_lc}). The signal at mid-transit displays a decrease of $\sim$50\% in flux in relation to the baseline, which is consistent with previous results obtained with \emph{HST}-STIS \citep{2015Natur.522..459E, 2017A&A...605L...7L}. These results indicate that the large atmospheric loss rate of GJ~436~b is stable on a timescale of a few years; however, it is difficult to evaluate the effects of stellar activity in the escape rate because of the large uncertainties and spread in the Lyman-$\alpha$ light curves obtained with COS. In addition to the uncertainties of the spectra calculated by the pipeline, the uncertainties shown in Fig. \ref{lya_lc} also include those of the airglow removal procedure; the latter were estimated by computing the flux for each airglow template from the MCMC simulation, and adding the spread in quadrature to the original uncertainty. In general, the uncertainties of the fit increase those of the resulting cleaned spectra by $\sim$10\%.

\subsection{Deep absorption event seen in the red wing}

\begin{figure}
\includegraphics[width=1.00\hsize]{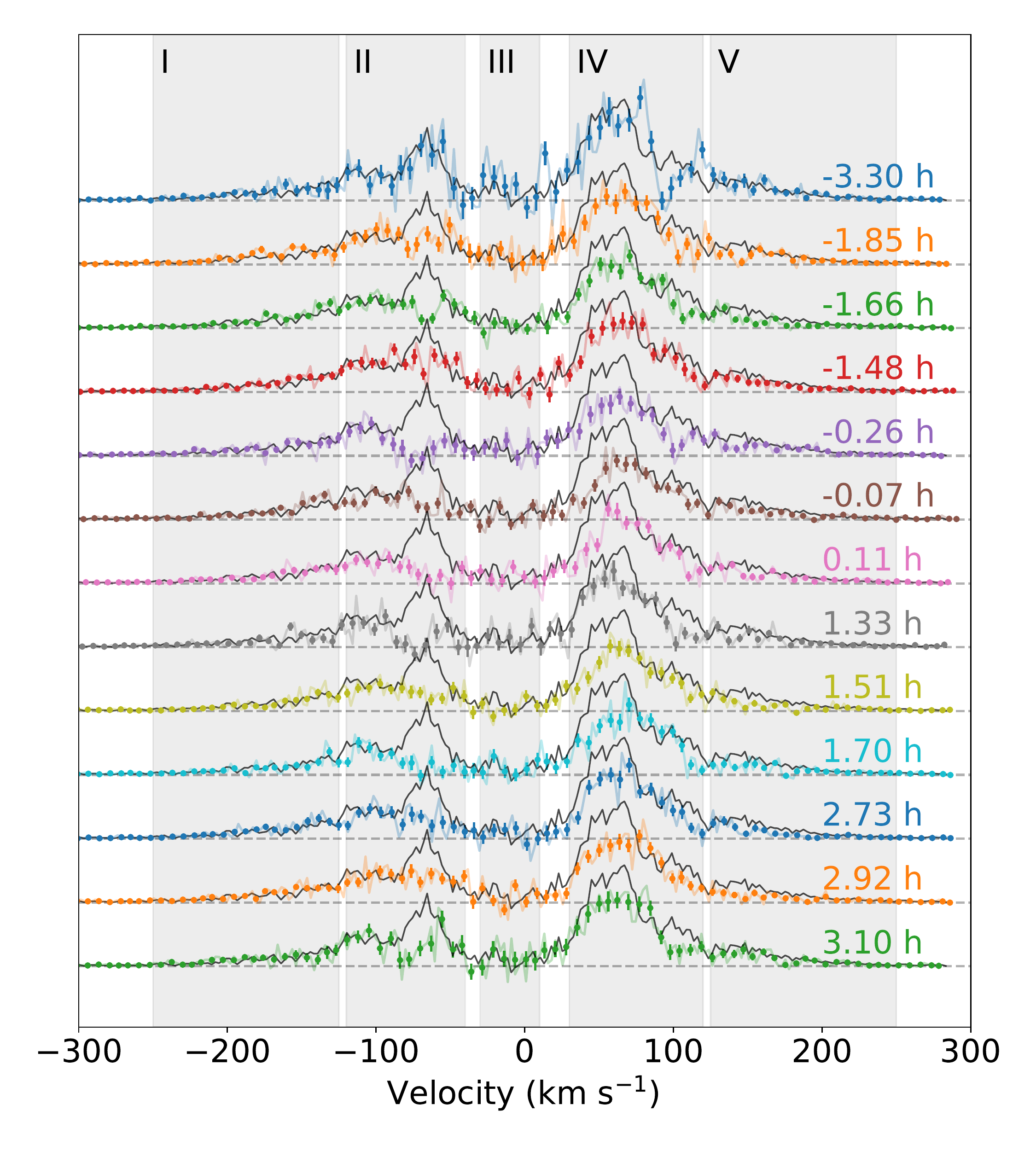}
\caption{Same as Fig. \ref{lya_pr}, but for subexposures of Visit C. In addition to the persistent in-transit absorption seen in the blue wing (region II), this visit also displays a 30\%-deep excess absorption signal in the Lyman-$\alpha$ red wing (region IV).}
\label{lya_visitC}
\end{figure}

\begin{figure}
\includegraphics[width=1.00\hsize]{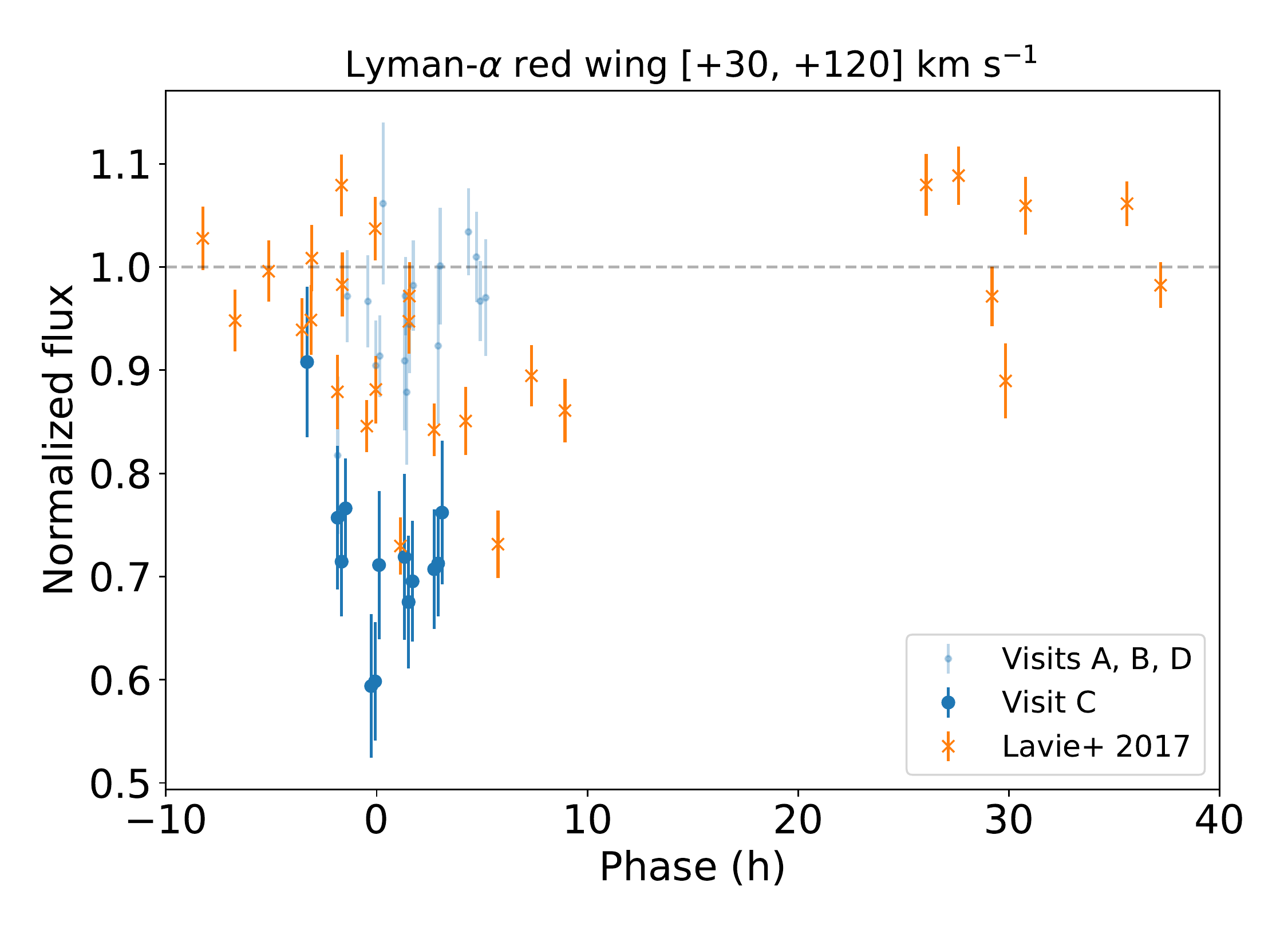}
\caption{Normalized light curve of red wing (region IV) fluxes during transit of GJ~436~b. The baseline flux used for normalization was computed in the same way as in Fig. \ref{lya_lc}. Visits A, B, and D display fluxes similar to the baseline value, while Visit C displays a decrease in flux by $\sim$30\% in relation to the baseline. The excess absorption in Visit C is likely physical and traces possible inflow of material to the host star.}
\label{lya_lc_red}
\end{figure}

We also analyzed the red wing of the Lyman-$\alpha$ line of GJ~436 in the wavelength region where \citet{2017A&A...605L...7L} had previously suggested an absorption signal at +5.75 h after mid-transit (region IV in Fig. \ref{lya_visitC}). We found that the observed Lyman-$\alpha$ red wing fluxes during Visit C are $\sim$30\% lower than the other visits and the out-of-transit exposures (see the time series in Figs. \ref{lya_visitC} and \ref{lya_lc_red}). We do not expect the intrinsic stellar Lyman-$\alpha$ to decrease by $\sim$30\% in flux by chance or stellar activity alone, and the uncertainties in the airglow removal procedure do not account for this lower resulting flux. Even though our observations with COS have a similar precision as the STIS data, they do not cover the phase +5.75 h. Therefore, we were unable to reproduce an absorption signal similar to what is seen with STIS. The COS and STIS signals seem to have a similar shape, but they are shifted in phase space and are deeper with COS.

This excess absorption seen in the red wing indicates the presence of \ion{H}{I} atoms inflowing to the host star at speeds varying from 30 to 120 km s$^{-1}$. As pointed out by \citet{2017A&A...605L...7L}, the exospheric model of GJ~436~b produced by \texttt{EVE} \citep{2013A&A...557A.124B, 2015A&A...582A..65B, 2016A&A...591A.121B} predicts that the population of \ion{H}{I} atoms moving toward the star is localized in the coma of the planet, producing potential signatures of up to 50 km~s$^{-1}$ only. Furthermore, \citet{2017A&A...605L...7L} suggests that star-planet interactions (SPIs) could explain redshifted signatures \citep[as in][]{2015A&A...578A...6M, 2016ApJ...833..140S}. It is unclear, however, how stable SPI signatures are in orbital phase space.

Similar redshifted in-transit excess absorption signals were marginally detected for HD~189733~b \citep{2012A&A...543L...4L} and HD~209458~b \citep{2003Natur.422..143V}. A persistent and significant signal was observed in the stellar Lyman-$\alpha$ red wing during the transit of GJ~3470~b \citep{2018A&A...620A.147B}. In order to explain this feature, \citeauthor{2018A&A...620A.147B} argues that the excess redshifted absorption is caused by the damping wings of dense layers of neutral hydrogen that extend beyond the planetary Roche lobe and are elongated in the direction of the orbital motion \citep[see, e.g.,][]{2005ApJ...621.1049T, 2008ApJ...688.1352B}. These observations do not constrain the line-of-sight position of this layer of \ion{H}{I} atoms, but \citet{2018A&A...620A.147B} suggest that they could be located in the shock interface between the planetary thermosphere and the stellar wind. A direct comparison with the episodic redshifted signal observed in GJ~436~b is not straightforward since the orbital configuration and systemic properties are different. However, detailed modeling of the interaction between the upper atmosphere of GJ~436~b and the stellar wind, particularly during and after flares, could provide an explanation for the observed redshifted signal.

\section{Searching for planet-induced variability signals in metallic lines}\label{results}

\citet{2017ApJ...834L..17L} reported on observations performed for the MUSCLES program and concluded that there was no absorption signal in the \ion{Si}{III} line and, with 95\% confidence, ruled out signals with depths larger than 49\% (in the case of a highly asymmetrical transit similar to what is observed in Lyman-$\alpha$). Furthermore, \citet{2017ApJ...834L..17L} reported a non-detection of \ion{C}{II} absorption in the transmission spectrum of GJ~436~b. \citet{2017A&A...605L...7L} obtained a tentative detection of absorption during transit in the \ion{Si}{III} line using \emph{HST}-STIS; they reported a transit depth of $47 \pm 10\%$ for the line flux in the interval [-50, 50] km s$^{-1}$. However, \citeauthor{2017A&A...605L...7L} also cautioned that they could not rule out possible stellar variability on the line, similarly for the red wing of the Lyman-$\alpha$ line. Aiming to reproduce and improve upon these previous results, we searched for possible exospheric absorption signals in the metallic FUV lines of GJ~436 in several datasets available to us.

\subsection{Excess in-transit absorption}\label{in-transit-abs}

\begin{figure*}
\centering
\begin{tabular}{cc}
\includegraphics[width=0.44\hsize]{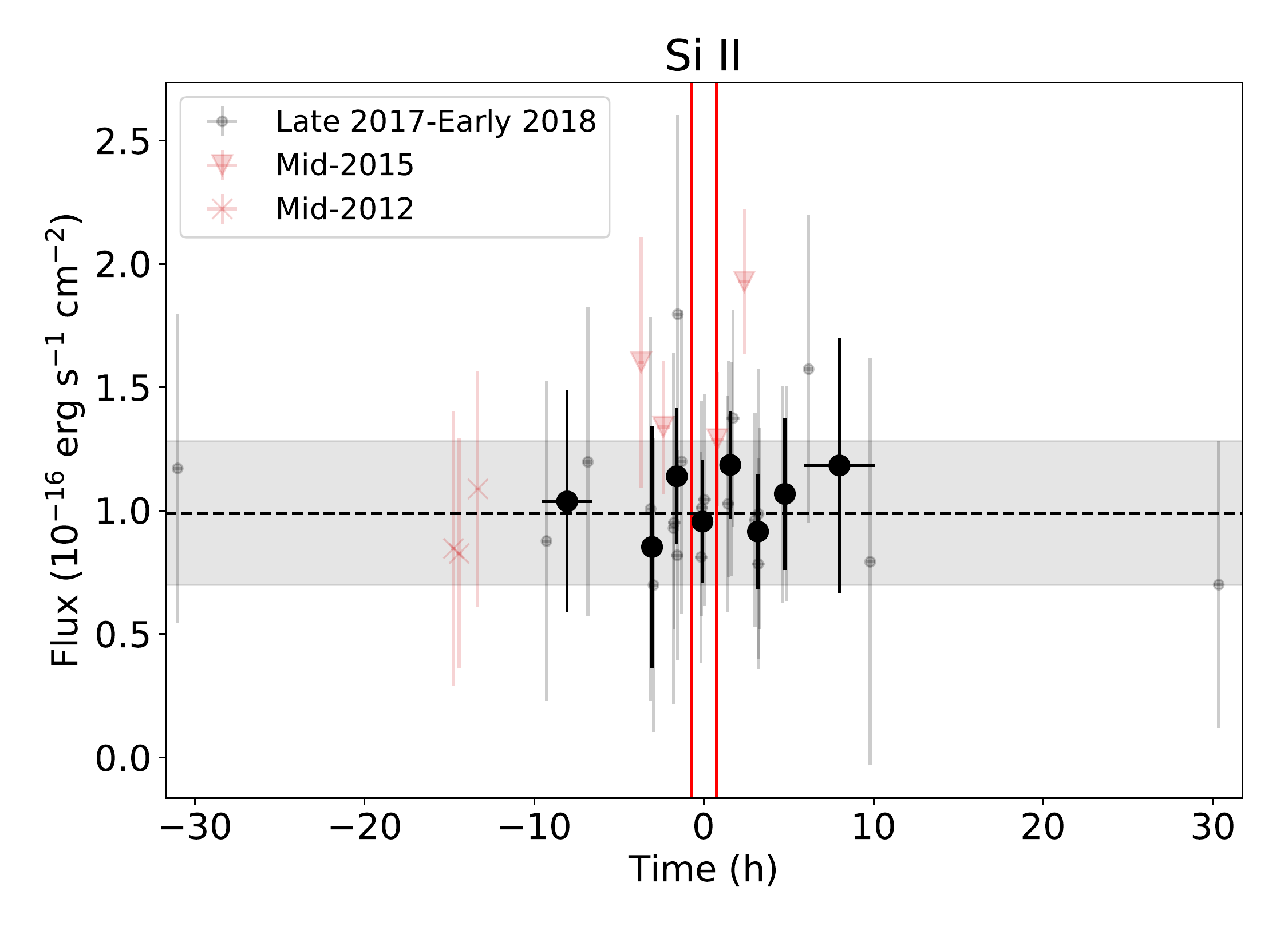} & \includegraphics[width=0.44\hsize]{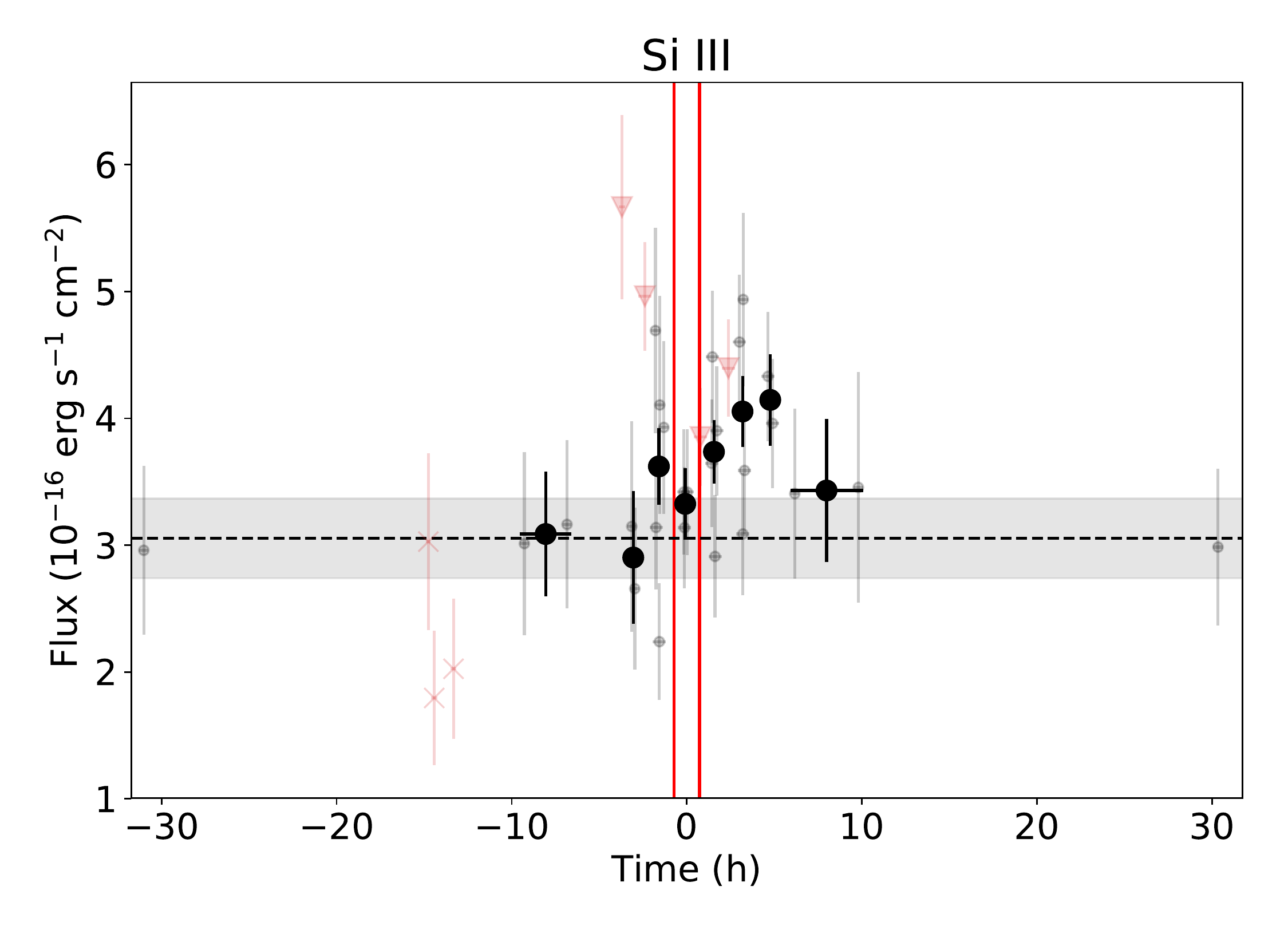} \\
\includegraphics[width=0.44\hsize]{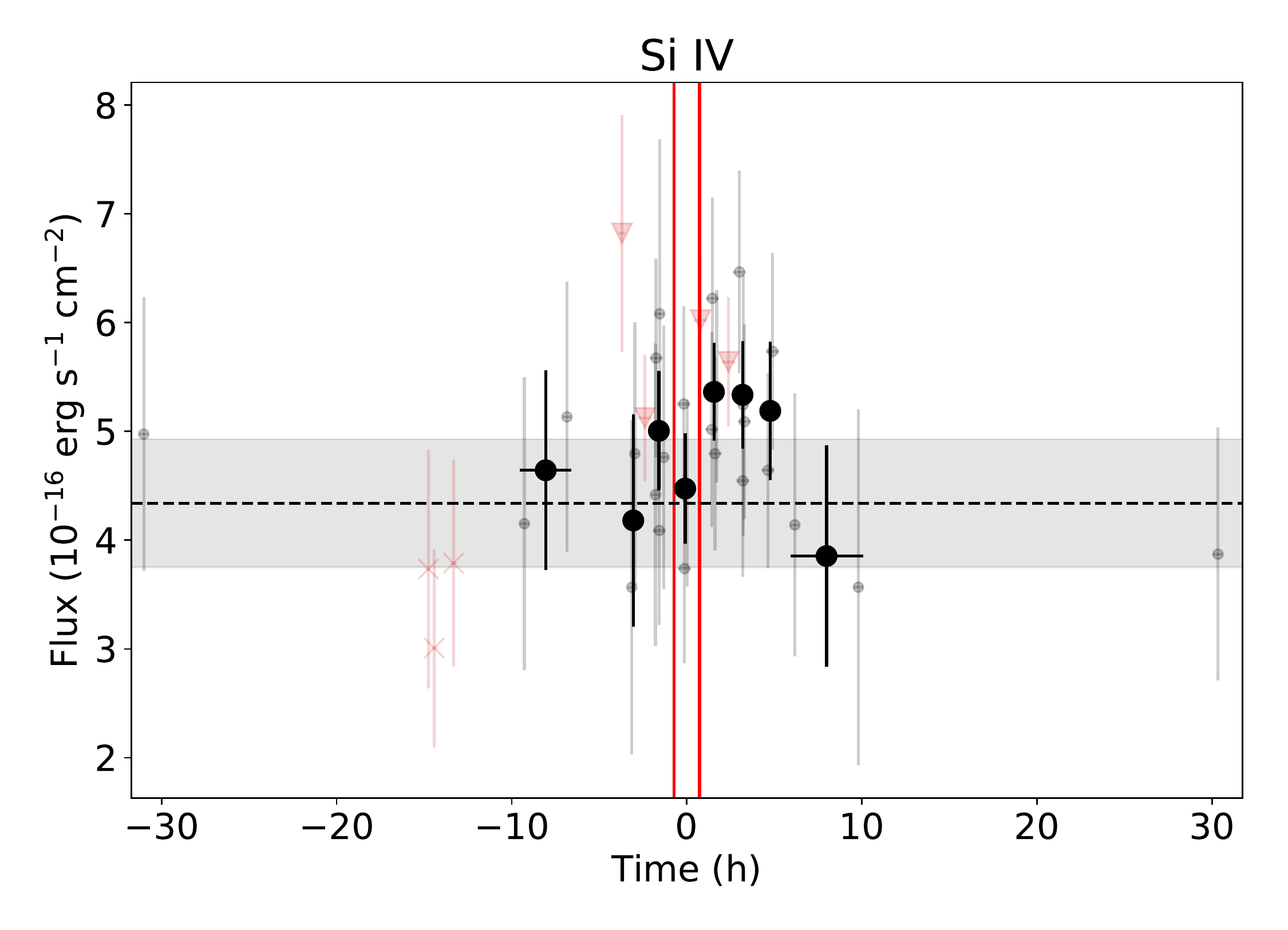} & \includegraphics[width=0.44\hsize]{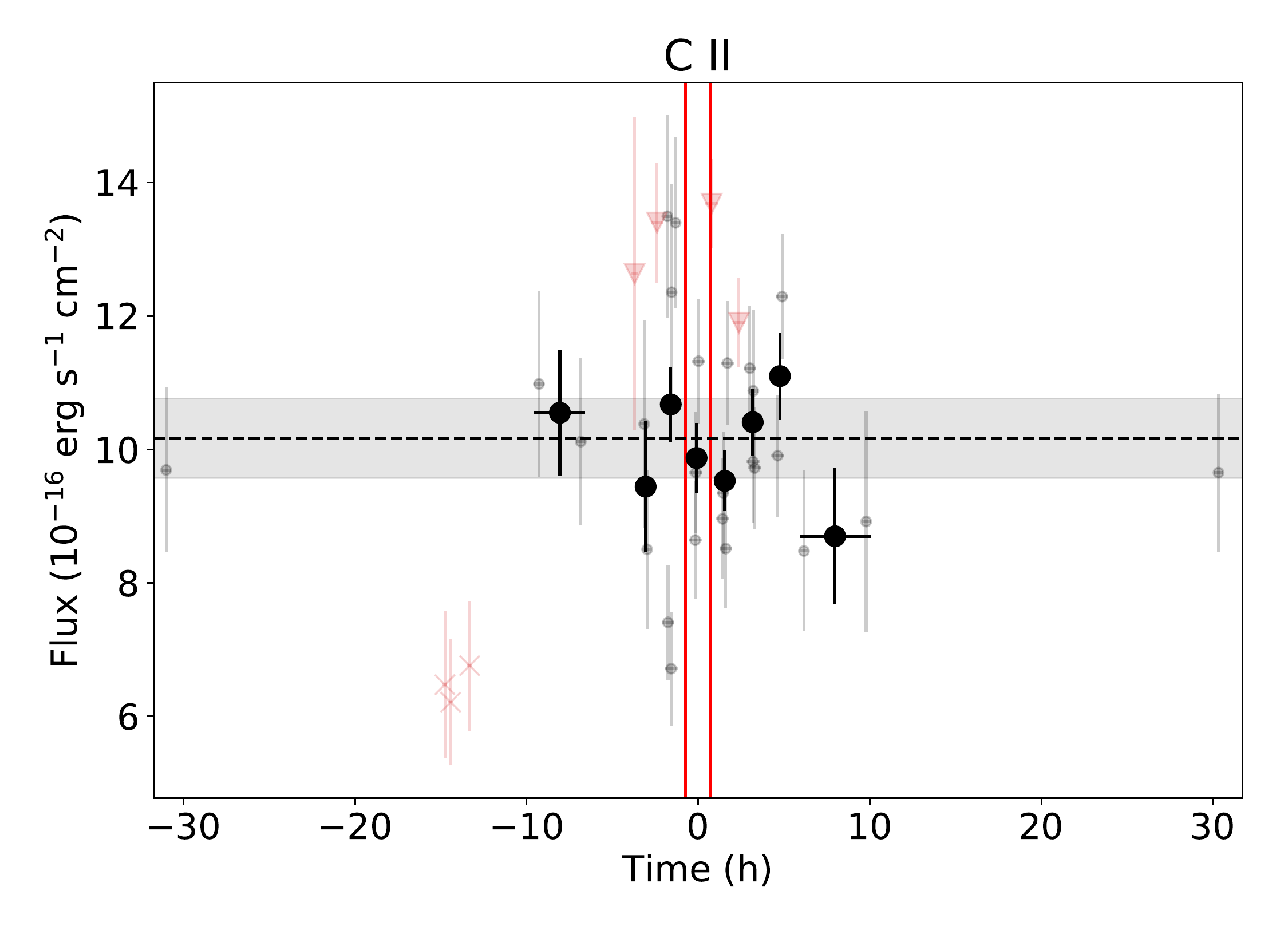} \\
\includegraphics[width=0.44\hsize]{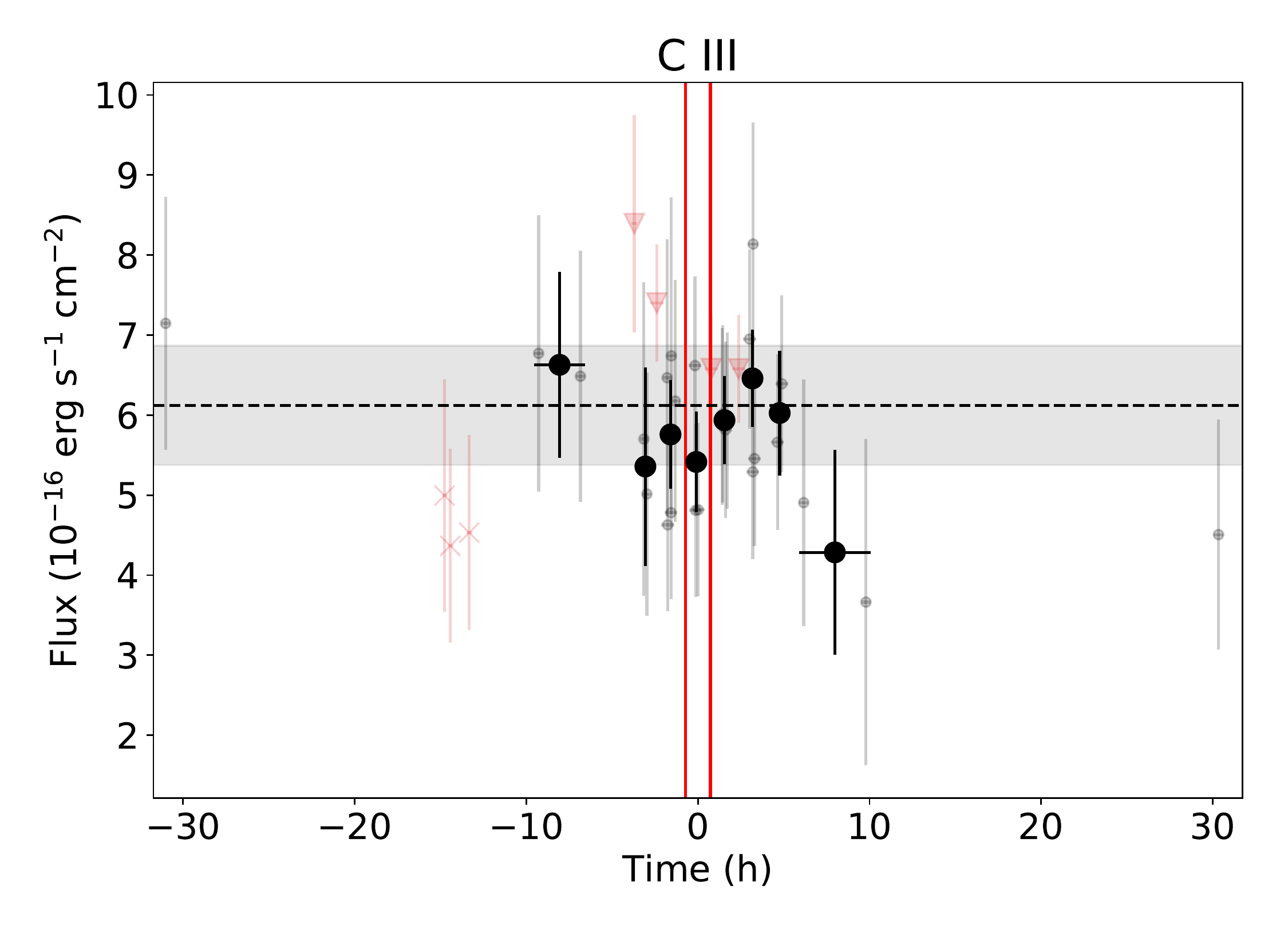} & \includegraphics[width=0.44\hsize]{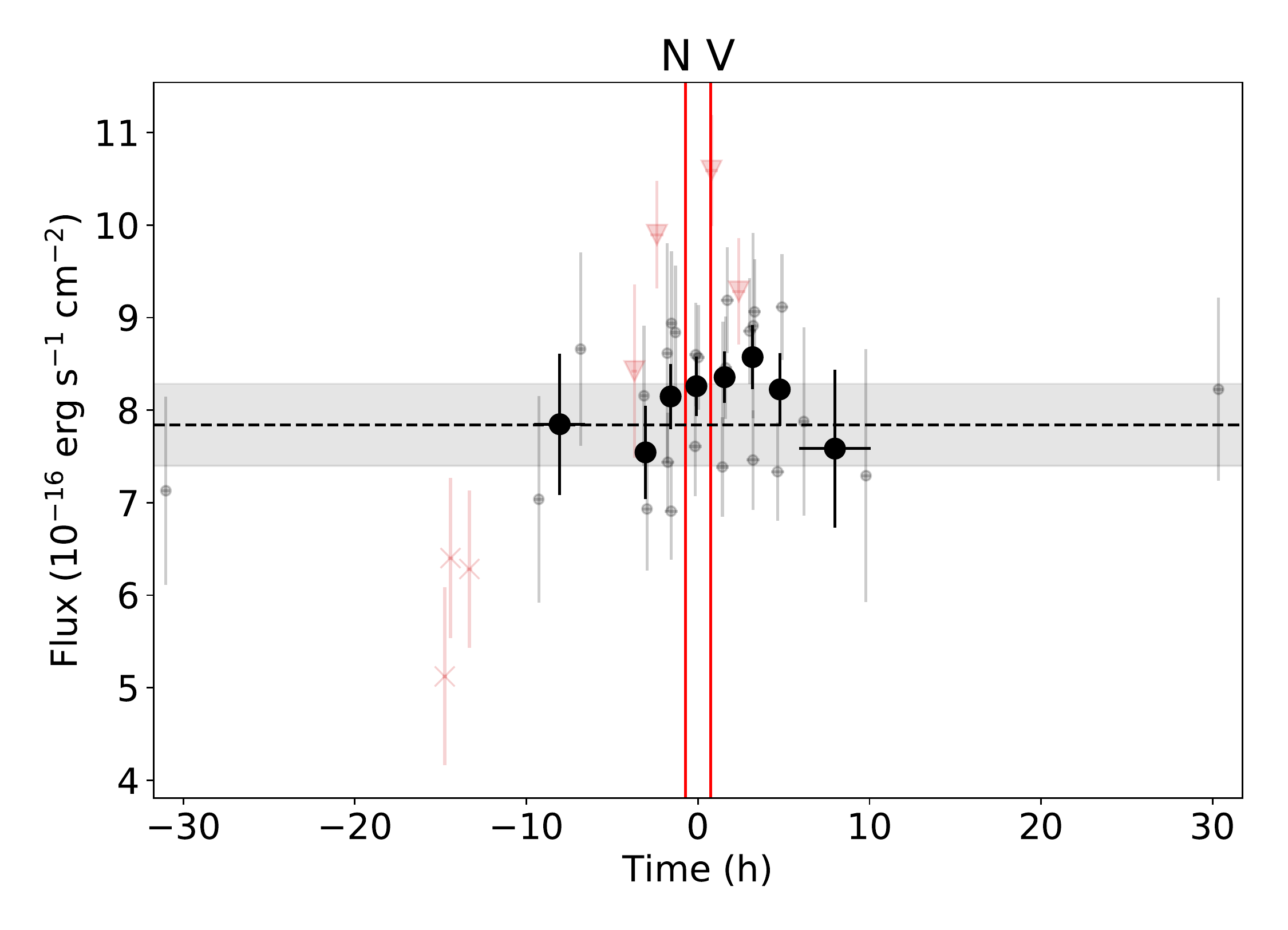} \\
\includegraphics[width=0.44\hsize]{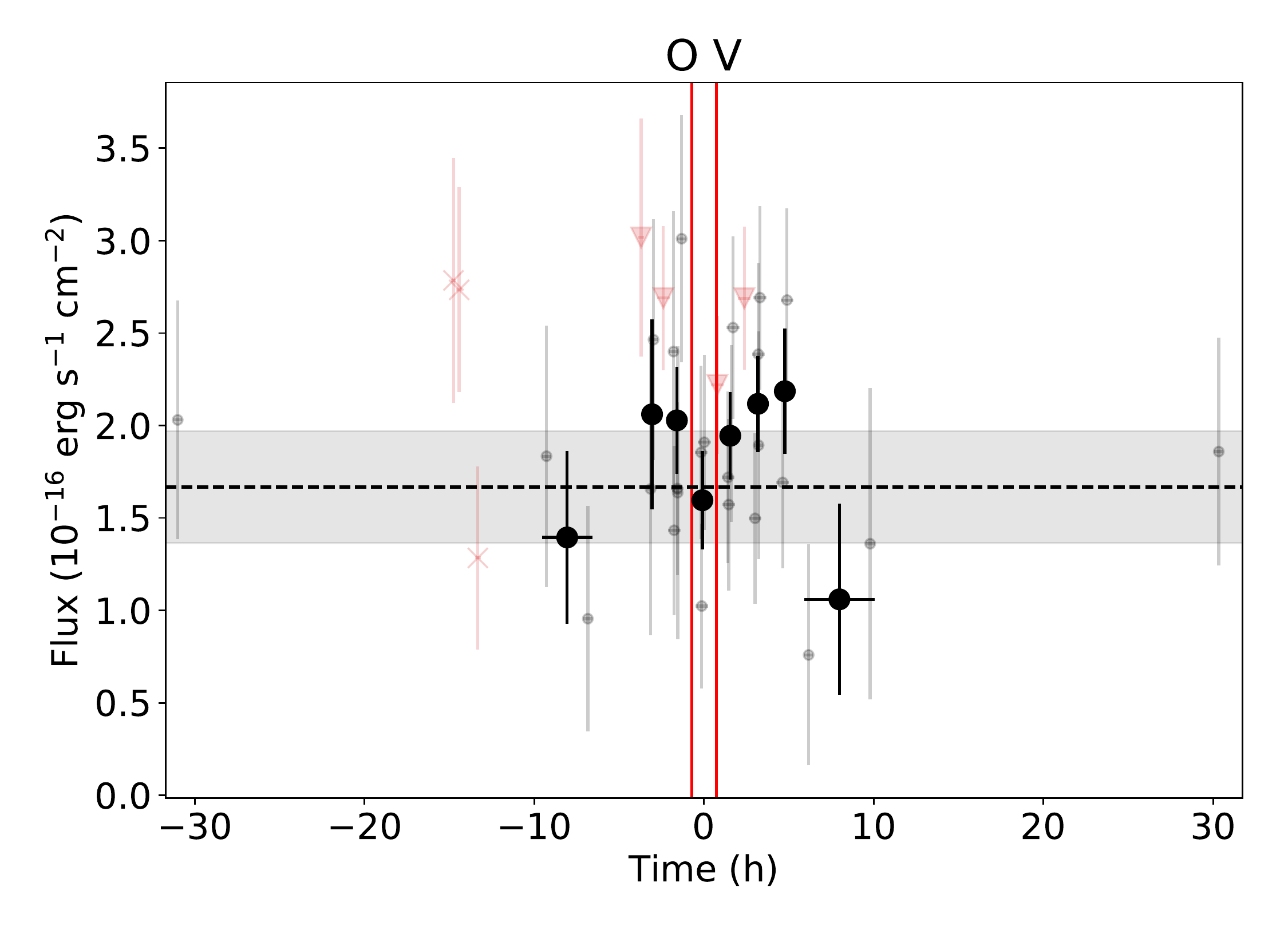} & \includegraphics[width=0.44\hsize]{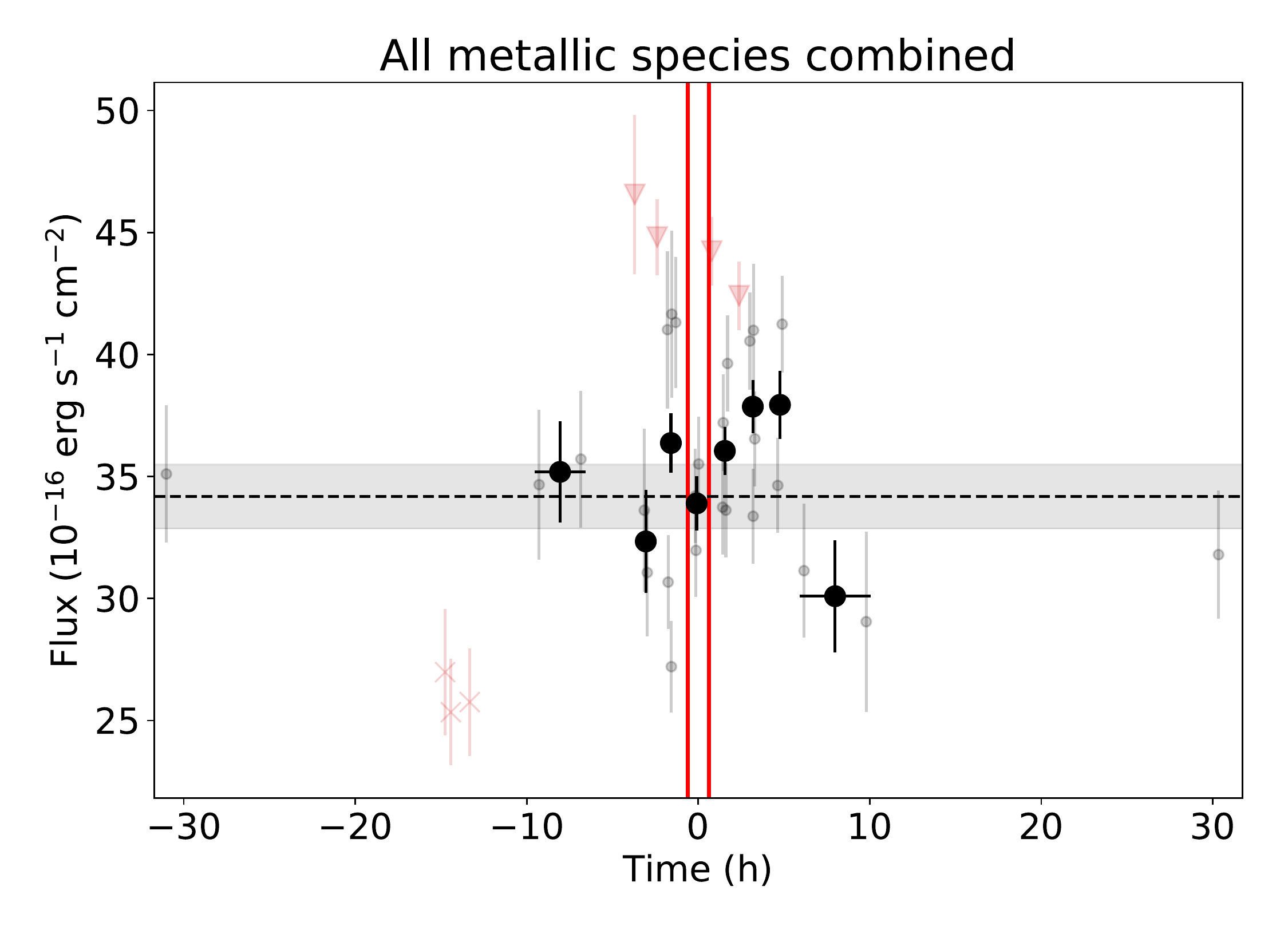} \\
\end{tabular}
\caption{Light curves of FUV stellar lines during transit of GJ~436~b. The larger circular symbols represent the late 2017-early 2018 data binned in phase. The red vertical lines represent the ingress and egress of the optical transit. The vertical dashed line and the gray region represent, respectively, the mean and 1$\sigma$ uncertainty of the baseline measured asymmetrically, similarly to the Lyman-$\alpha$ baseline.}
\label{lc_full}
\end{figure*}

In order to increase the signal-to-noise ratios of the phase-folded light curves, we binned the fluxes in phase space (black circles in Fig. \ref{lc_full}). We included the visits from programs previous to the 2017-2018 epoch in our plots for comparison purposes only, but they are not to be taken into account when computing baseline fluxes and detection levels. As discussed in Sect. \ref{rot_modulation_sect}, the exposures taken during the epochs in 2012 and 2015 may correspond to different phases of the magnetic cycle of GJ~436, so it is difficult to correct for activity effects in them without continuously monitoring them. Thus, we did not apply rotational modulation correction to the 2017-2018 epoch since the effect is not significant in this period alone (it is only significant if we assume that the activity modulation in mid-2015 is coherent with the most recent epoch).

It is not straightforward to interpret the flux time series for each species due to the strong variability. If we fix the confidence level at 95\% (which corresponds to 2$\sigma$), we could rule out different levels of mid-transit absorption signals depending on the precision with which we can measure the baseline stellar flux for each species. The baseline flux itself is uncertain for the metallic lines; we adopted the following three different definitions of baseline for the purpose of determining the non-detection levels of absorption: i) one similar to the optical transit (based on the parameters in Table \ref{gj436_param}) in which the baseline is measured in phases outside the optical transit; ii) a long symmetric transit, for which the baseline is measured with every data point outside the phase range [-5, 5] h; and iii) an asymmetric transit, for which the baseline is measured in the same way as the Lyman-$\alpha$ flux, namely outside the phase range [-3, +24] h. These non-detection levels are summarized in Table \ref{results_transit}, and they represent the minimum levels of excess absorption or emission that can be detected if such signals were present in the data.

\begin{table}
\caption{Minimum levels of planet-induced variability signals in metallic lines of GJ~436 that we can rule out at 95\% confidence (2$\sigma$).}
\label{results_transit}
\centering
\begin{tabular}{lccc}
\hline\hline
\multirow{2}{*}{Species} & \multicolumn{3}{c}{Absorption depth} \\
& (opt. transit) & (long transit) & (asym. transit)\\
\hline
\ion{C}{II} & 4.8\% & 11\% & 12\% \\
\ion{C}{III} & 10\% & 24\% & 24\% \\
\ion{Si}{II} & 23\% & 51\% & 59\% \\
\ion{Si}{III} & 7.4\% & 18\% & 21\% \\
\ion{Si}{IV} & 9.8\% & 25\% & 27\% \\
\ion{N}{V} & 4.2\% & 12\% & 11\% \\
\ion{O}{V} & 7.8\% & 20\% & 20\% \\
\hline
\end{tabular}
\tablefoot{The minimum detectable signals of variability depend on how the baseline is defined. The different baseline definitions we adopted are outlined in Section \ref{in-transit-abs}.}
\end{table}

\begin{figure}
\centering
\includegraphics[width=\hsize]{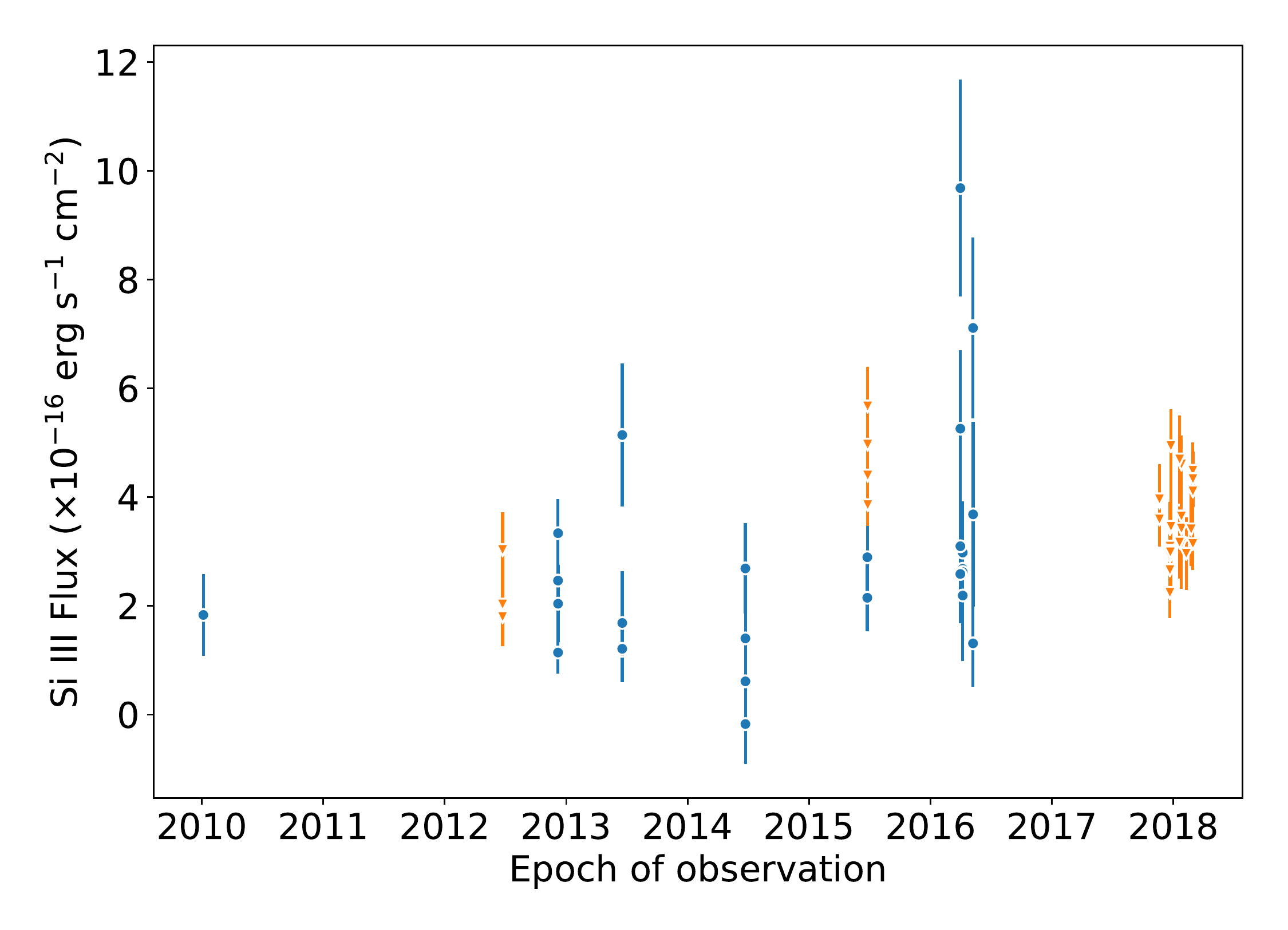}
\caption{Stellar \ion{Si}{III} fluxes of GJ~436 measured with STIS and COS spectrographs on \emph{HST} across different epochs. The average flux and variability of the line seem to have increased since 2015, which can be related to the stellar activity cycle.}
\label{SiIII_epochs}
\end{figure}

Since we were unable to reproduce the result over several visits with \emph{HST}-COS, it is likely that the \ion{Si}{III} absorption signal reported by \citet{2017A&A...605L...7L} is related to stellar variability instead of absorption by the exosphere of GJ~436~b. As discussed in Sect. \ref{rot_modulation_sect}, approximately half of the STIS observations were obtained in 2016 when the star was coming out of an activity maximum and the \ion{Si}{III} fluxes, which are very sensitive to stellar activity, increased in variability and average flux (see Fig. \ref{SiIII_epochs}). The previous observations were obtained between 2013 and 2015, the epochs when GJ~436 was coming out of an activity minimum. The STIS observations after 2015 were specifically performed to cover orbital phase ranges farther from mid-transit, so that explains the higher \ion{Si}{III} baseline inferred by \citet{2017A&A...605L...7L} in relation to the exposures near mid-transit, which were executed before the increase in activity.

Our results do not, however, rule out the 2\% absorption depth in the \ion{C}{II} lines predicted by \citet{2017ApJ...834L..17L}, and more observations would in principle be necessary to confirm their prediction. Our analysis indicates that the \ion{C}{II} lines are particularly sensitive to stellar activity, so the detection of such a shallow signal may be very challenging using the current FUV instrumentation. In the future, more sensitive instruments, such as the Large Ultraviolet/Optical/Infrared Surveyor \citep[\emph{LUVOIR;}][]{2017SPIE10398E..09B}, the LUVOIR Ultraviolet Multi-Object Spectrograph \citep[LUMOS;][]{2017SPIE10397E..13F}, and the Habitable Planet Explorer \citep[\emph{HabEx};][]{2016SPIE.9904E..0LM} will be able to measure FUV fluxes with several times better precision than \emph{HST}-COS.

Neptune-sized planets have a lower surface gravity than Jupiter-sized ones, so in principle we would expect the first to lose heavier atoms more easily than the second. However, it is likely that these atoms condense into clouds more easily in warm Neptunes and, therefore, they cannot be carried upward to the exosphere as easily \citep{2017ApJ...834L..17L}. In the particular case of GJ~436~b, the planet is relatively cool \citep[$T_\mathrm{eq} \sim 600$ K;][]{2016MNRAS.459..789T} when compared to hot Jupiters ($T_\mathrm{eq} > 1000$ K) where heavier elements have been detected in their extended atmospheres. The upper limits of absorption levels of metallic ions during the transit of GJ~436~b are similar to the absorption signals detected for the hot Jupiters HD~209458~b \citep{2004ApJ...604L..69V} and HD~189733~b \citep{2013A&A...553A..52B}. Our results could indicate that either: first, the escape process in GJ~436~b is not hydrodynamic, but hydrostatic; or second, by assuming hydrodynamic escape, mixing in the lower atmosphere is not efficient in dragging the metal-rich clouds high enough for sublimation and allow for a significant escape rate of metallic ions.

\section{Conclusions}\label{conclusions}

We reported on the analysis of \emph{HST}-COS observations of the stellar FUV spectra during four transits of the planet GJ~436~b obtained for the PanCET program, as well as archival \emph{HST} data from the MUSCLES and GO-15174 programs. Even though GJ~436 is considered a quiet M dwarf when compared to other similar stars, it displays flaring activity such as the events reported by \citet{2017ApJ...834L..17L}. Our analysis revealed that GJ~436 also displays flare activity that increases the fluxes of \ion{C}{II} and \ion{Si}{III} by $\sim$50\% and $\sim$200\%, respectively, and returns to quiescent levels in 20 minutes or less -- such a behavior is also observed in X-ray light curves of the Sun and GJ~436 (Sanz-Forcada et al., in prep.). In total, we found seven events with flux brightenings in the FUV spectra of GJ~436 in the PanCET and the archival data, resulting in a flaring rate of 10.1 d$^{-1}$. Some of these brightening events are not apparent in other FUV spectral lines that are less sensitive to activity, such as Lyman-$\alpha$.

The FUV fluxes of GJ~436 taken in the 2017-2018 epoch do not display significant rotational modulation if analyzed alone. However, if we assume that the mid-2015 traces the same, long-lived activity region or the same active longitude, then the strongest metallic lines, namely \ion{C}{II}, \ion{Si}{III,} and \ion{N}{V} display significant rotational modulation. In the latter case, the amplitudes of the modulation would be approximately 20\% for the first two and 10\% for the last, and they would all appear to be in phase. Slowly-rotating M dwarfs similar to GJ~436 have long-lived activity regions that can last for many years \citep[such as GJ~176 and Proxima Centauri;][]{2015ApJ...801...79R, 2016A&A...595A..12S}. Analysis of the STIS data also suggests a marginal rotational modulation of the Lyman-$\alpha$ line of GJ~436 with a $\sim$10\% amplitude in phase with the COS fluxes of the stellar metallic lines. Future observations of stars like GJ 436 in short wavelengths require carefully planned monitoring to cover the entire rotational phase of the star in order to remove effects of rotational modulation and variability.

The \emph{HST}-COS observations centered at 1291 \AA\ include the stellar Lyman-$\alpha$ emission, but it is severely contaminated by the Earth's geocoronal emission in comparison to STIS exposures because COS possesses a circular aperture. Removing the airglow contamination in this case is not a trivial process, especially when the stellar emission is fainter than the airglow, which is the case for GJ~436. We performed the same Lyman-$\alpha$ correction procedure as in \citet{2018A&A...615A.117B} to estimate the stellar emission from the COS observations and to recover the stellar Lyman-$\alpha$ emission for the subexposures that were performed near the Earth's shadow (namely when the airglow and stellar emission levels were comparable).

We searched for potential atmospheric signals caused by the planet GJ~436~b transiting its host star. We were able to reproduce the Lyman-$\alpha$ blue wing light curve during the transit of GJ~436~b that had previously revealed that the planet possesses a large exosphere that produces a $\sim$50\% decrease in the stellar emission between [-120, -40] km s$^{-1}$ in Doppler velocities. We conclude that the excess absorption in the Lyman-$\alpha$ blue wing is stable for several years. In addition, one of the PanCET visits, more specifically the one obtained in late January 2018, displays a significant excess absorption of $\sim$30\% in the Lyman-$\alpha$ red wing (between [+30, +120] km s$^{-1}$ in Doppler velocities). This potential in-transit signal in the red wing occurs during the whole visit, and it is deeper and shifted in phase when compared to the ones reported by \citet{2017A&A...605L...7L}. However, it is not reproduced in the other PanCET visits, indicating a temporary and possibly stochastic event.

Several metallic lines of ions in the transition region of GJ~436 are present in our datasets, with the brightest being \ion{C}{II}, \ion{C}{III}, \ion{Si}{II}, \ion{Si}{III}, \ion{Si}{IV}, \ion{N}{V,} and \ion{O}{V}. The in-transit light curves of the combined fluxes for each species do not reveal any evidence that leads us to conclude that such ions are present in the exosphere of the planet. In particular, assuming an asymmetrical transit similar to Lyman-$\alpha$, we can rule out an absorption depth of 12\% and 11\% for the \ion{C}{II} and \ion{N}{V} fluxes, respectively, with 95\% confidence during the transit of GJ~436~b, which is consistent with the results of \citet{2017ApJ...834L..17L}. On the other hand, we were not able to reproduce the in-transit absorption signal in \ion{Si}{III} that was suggested by \citep{2017A&A...605L...7L}; it is likely that this signal was caused by the increased \ion{Si}{III} fluxes of GJ~436 after 2015, which is when the observations of the baseline flux were made and coincidentally the star was coming out of a maximum in its activity cycle. A large observational effort may be necessary to put stricter constraints on the presence of Si ions in the upper atmosphere GJ~436~b.

We are still trying to better understand the atmosphere of GJ~436~b. The FUV transmission spectrum gives us access to the upper atmopshere, while optical and infrared spectra trace the lower atmopshere. Given its featureless optical transmission spectrum, it is still not completely clear if it has a high metallicity or a cloudy atmosphere \citep{2018AJ....155...66L}. Using \emph{Spitzer} photometry at 3.6, 4.5, and 8 $\mu$m, \citet{2014A&A...572A..73L} concluded that their results are consistent with a metal-rich atmosphere depleted in methane and enhanced in CO and CO$_2$. The non-detection of metallic species in its exosphere, in particular Si, suggests that if GJ~436~b possesses a cloudy atmosphere and if the escape is hydrodynamic, then mixing is not efficient in dragging the Si-rich clouds high enough for sublimation which would allow for a significant escape rate of metallic ions. On the other hand, non-detection cannot rule out a hydrostatic escape process instead.

\begin{acknowledgements}
LAdS warmly thanks A. Su\'arez Mascare\~no, H. Cegla, H. Giles, J. Seidel, N. Hara, and K. France for the fruitful discussions about stellar activity and modeling during the preparation of this manuscript. We are also grateful for the valuable feedback from the anonymous referee and support provided by the Help Desk of STScI. This project has received funding from the European Research Council (ERC) under the European Union’s Horizon 2020 research and innovation programme (project {\sc Four Aces}; grant agreement No 724427), and it has been carried out in the frame of the National Centre for Competence in Research PlanetS supported by the Swiss National Science Foundation (SNSF). This research is based on observations made with the NASA/ESA Hubble Space Telescope. The data was obtained for the Hubble Panchromatic Comparative Exoplanet Treasury program and is openly available in the Mikulski Archive for Space Telescopes (MAST), which is maintained by the Space Telescope Science Institute (STScI). STScI is operated by the Association of Universities for Research in Astronomy, Inc. under NASA contract NAS 5-26555. This research made use of the NASA Exoplanet Archive, which is operated by the California Institute of Technology, under contract with the National Aeronautics and Space Administration under the Exoplanet Exploration Program. We used the open source software SciPy \citep{scipy_ref}, Jupyter \citep{Kluyver:2016aa}, Astropy \citep{2013A&A...558A..33A}, Matplotlib \citep{Hunter:2007}, \texttt{emcee} \citep{2013PASP..125..306F} and \texttt{batman} \citep{2015PASP..127.1161K}.
\end{acknowledgements}

\bibliographystyle{aa}
\bibliography{biblio.bib}

\end{document}